\documentclass[10pt,twocolumn,letterpaper]{article}

\usepackage[pagenumbers]{iccv} 

\definecolor{tabfirst}{rgb}{1, 0.7, 0.7}
\definecolor{tabsecond}{rgb}{1, 0.85, 0.7}
\definecolor{tabthird}{rgb}{1, 1, 0.7}
\definecolor{tabgray}{rgb}{0.9, 0.9, 0.9}

\usepackage[accsupp]{axessibility}  
\definecolor{iccvblue}{rgb}{0.21,0.49,0.74}
\usepackage[pagebackref,breaklinks,colorlinks,allcolors=iccvblue]{hyperref}
\usepackage{colortbl,xcolor}
\newif\ifdraft
\drafttrue 
\ifdraft
\newcommand{\NGC}[1]{{\color{red}[\textbf{Nithin:} #1]}}
\newcommand{\SKC}[1]{{\color{blue}[\textbf{Srinivas:} #1]}}

\else
\newcommand{\NGC}[1]{}
\newcommand{\SKC}[1]{}
\fi

\def\abstract{%
   \centerline{\large\bf Abstract}%
   \vspace*{6pt}\noindent%
   \it\ignorespaces%
}

\def\paperID{3642} 
\def\confName{ICCV}
\def\confYear{2025}
\title{Scaling  Transformer-Based Novel View Synthesis Models with \\ Token Disentanglement and Synthetic Data}
\author{
Nithin Gopalakrishnan Nair$^{1*}$\quad Srinivas Kaza$^{2*}$\quad Xuan Luo$^{2}$\quad Vishal M. Patel$^{1}$\quad \\ Stephen Lombardi$^{2}$\quad  Jungyeon Park$^{2}$\quad 
\\
\small{
$^{1}$~Johns Hopkins University\quad
$^{2}$~Google
}
\\
{\scriptsize \texttt{\{ngopala2,vpatel36\}@jhu.edu} \quad \texttt{\{srinivaskaza,xuluo,salombardi,jungyeonp\}@google.com}} \\
\small{\url{https://scaling3dnvs.github.io}}
\vspace{-0.21cm}
}
\begin{document}
\twocolumn[{%
\renewcommand\twocolumn[1][]{#1}%
\maketitle
\vspace{1.5cm}
\vspace{-3\baselineskip}
\vspace{-3\baselineskip}
\begin{center}
\centering
\setlength{\tabcolsep}{0.5pt}
\captionsetup{type=figure}
{\footnotesize
\renewcommand{\arraystretch}{0.5} 
\begin{tabular}{c}
     \includegraphics[width=\linewidth]{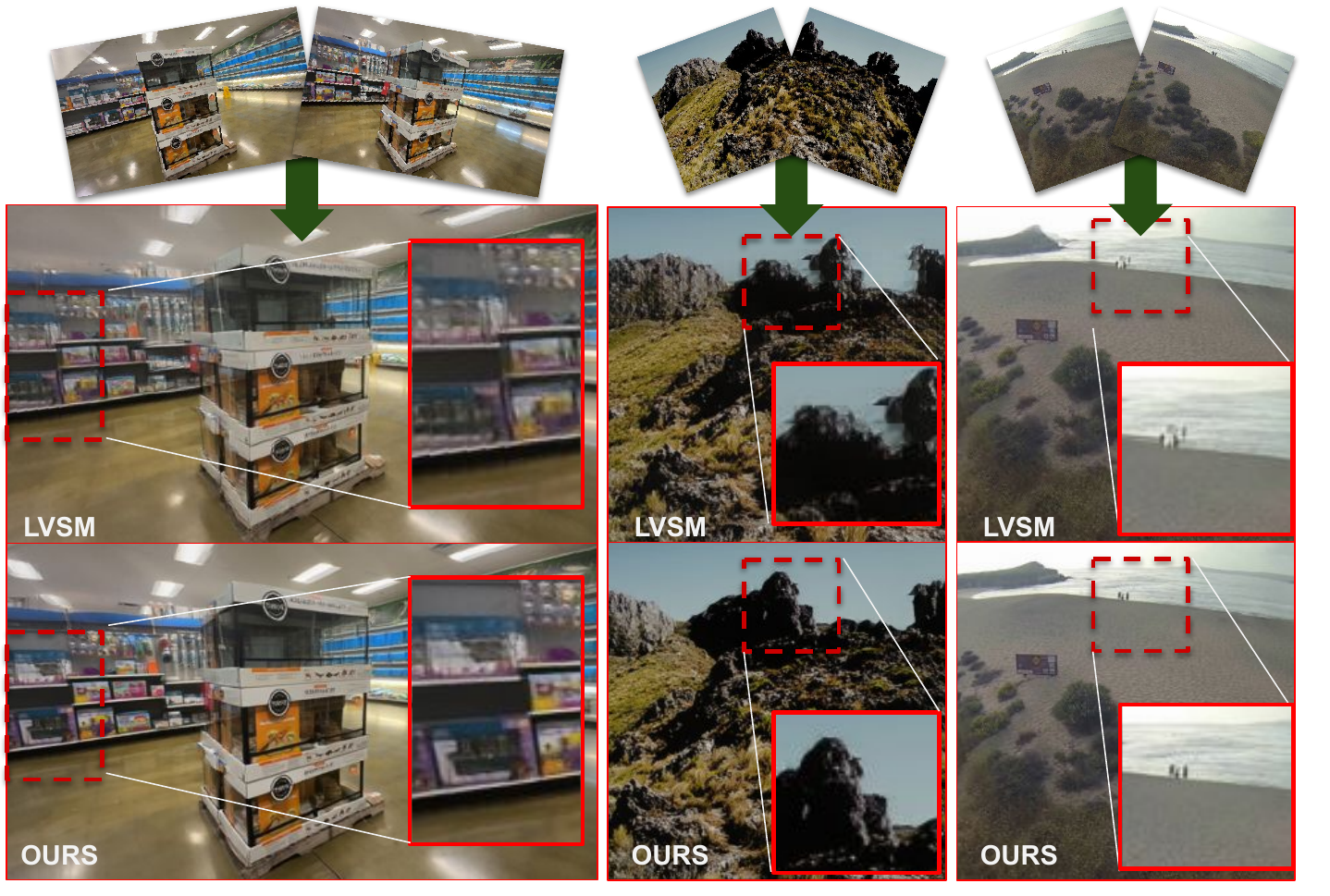}
\tabularnewline
\end{tabular}}
\vspace{-0.5\baselineskip}
\vspace{-3mm}
\hspace{20pt}\captionof{figure}{\textbf{Overview.} Our method performs feed-forward novel-view synthesis from a series of input images, such as the pairs shown above. We demonstrate strong results in terms of quality and generalization capacity, performing well across a variety of common novel-view synthesis datasets, including scenes that are out-of-distribution.}
\label{fig:introfig}
\end{center}%
}]
\thispagestyle{empty}
\vspace{-13mm}
\begin{abstract}
Large transformer-based models have made significant progress in generalizable novel view synthesis (NVS) from sparse input views, generating novel viewpoints without the need for test-time optimization. However, these models are constrained by the limited diversity of publicly available scene datasets, making most real-world (in-the-wild) scenes out-of-distribution. To overcome this, we incorporate synthetic training data generated from diffusion models, which improves generalization across unseen domains. While synthetic data offers scalability, we identify artifacts introduced during data generation as a key bottleneck affecting reconstruction quality. To address this, we propose a token disentanglement process within the transformer architecture, enhancing feature separation and ensuring more effective learning. This refinement not only improves reconstruction quality over standard transformers but also enables scalable training with synthetic data. As a result, our method outperforms existing models on both in-dataset and cross-dataset evaluations, achieving state-of-the-art results across multiple benchmarks while significantly reducing computational costs.
\renewcommand{\thefootnote}{\fnsymbol{footnote}}
\footnotetext[1]{Equal contribution. Nair designed the methodology, conducted pre-rebuttal experiments, and drafted the initial manuscript. Kaza helped advise the project, led the rebuttal, and conducted camera-ready experiments.}
\end{abstract}    
\vspace{-5mm}
\section{Introduction}
\label{sec:intro}
Novel view synthesis (NVS) \cite{mildenhall2020nerf, kerbl3Dgaussians} is a well-studied and important problem in computer vision, where the task is to generate unseen perspectives of a scene from a given set of images.  Many approaches utilize volumetric \cite{mildenhall2020nerf,barron2023zip,Chen2022ECCV,mueller2022instant} or differentiable rendering \cite{kerbl3Dgaussians} to optimize for each scene individually, achieving high-quality NVS from arbitrary viewpoints. More recently, advancements have enabled training a single model that generalizes to novel scenes without requiring per-scene optimization. Most existing methods address NVS by incorporating hand-crafted 3D priors and architectural biases \cite{suhail2022generalizable,hong2024lrm,charatan2024pixelsplat}. While these design choices provide structure, they limit scalability with data and hinder generalization.

Recently, Large View Synthesis Model (LVSM) \cite{jin2024lvsmlargeviewsynthesis} proposed a promising foundation for an NVS model scalable with large datasets. LVSM  introduces an architecture that doesn't require 3D inductive biases for scene reconstruction. It employs a decoder-only transformer architecture that achieves state-of-the-art results by a significant margin, with the performance improving with increased compute. However, we observed during our experiments that the decoder-only design causes an inherent feature alignment problem which causes the target and source features to look similar at all layers. Thus, part of the transformer's computational capacity is spent modifying source token information that is ultimately discarded at the end of the transformer block, reducing efficiency. This design choice also makes LVSM susceptible to unwanted noise or compression artifacts that may be present in the source views. In addition, we noticed that LVSM presents limited cross-domain performance when tested on datasets outside the training dataset domains.


Moreover, these issues are not unique to LVSM; many NVS models face similar challenges due to data scarcity in 3D vision. All existing multi-view 3D scene datasets \cite{Zhou2018,ling2024dl3dv,infinite_nature_2020} combined contain fewer than 100,000 scenes, severely limiting the performance of NVS models on in-the-wild cases beyond the training distribution. One possible solution for alleviating this 3D data scarcity is using synthetic data from generative models. Recent research has explored adapting pre-trained image \cite{saharia2022photorealistic,rombach2022high} and video diffusion models \cite{ho2020denoising,he2022latent} for multi-view dataset generation \cite{shi2023mvdream,xu2024depthsplat,gao2024cat3d, liu2020neural}. 
However, previous feed-forward models trained using synthetic data perform worse than those trained with real data. We hypothesize that the inability of synthetic data to improve reconstruction quality stems from two types of degradation artifacts in scenes generated by diffusion models \cite{ho2020denoising,nichol2021improved,song2020score} (1) artifacts influenced by the initial noise of the diffusion process and (2) artifacts introduced during decoding, as most diffusion-based scene synthesis models operate in latent space and rely on a diffusion VAE \cite{rombach2022high}.  We address both issues, leading to improved performance when using synthetic data. We provide a detailed explanation of our data pipeline in Section \ref{sec:syntheticpipline}.


In this work, we tackle a key challenge in developing a feed-forward NVS model that performs well on out-of-distribution data -- the need for a scalable and efficient transformer-based NVS architecture. We introduce the Token Distentangled (Tok-D) transformer block, which applies layer-wise modulation of source and target tokens, explicitly distinguishing the two at each layer. These model modifications improve out-of-distribution training, which introduces the possibility of training on synthetic data. We use the CAT3D model to generate a large dataset of synthetic multi-view samples. We then employ a novel data generation strategy that significantly improves the quality of these synthetic samples. We show that the Tok-D transformer block can be trained with synthetic data augmentation, unlike the baseline LVSM method which suffers from the inclusion of synthetic data.

\begin{itemize}[noitemsep]
    \item  We enhance the scalability of transformer architectures for NVS, enabling more efficient modeling.
    \item  We introduce a new training scheme that is less susceptible to artifacts from synthetic data.
    \item  We improve the training efficiency of transformer for NVS by introducing a new transformer block.
    \item Our approach achieves state-of-the-art results across multiple benchmarks for scene level NVS.
\end{itemize}

\section{Related Works}
 \begin{figure*}
    \centering
    \includegraphics[width=0.99\linewidth]{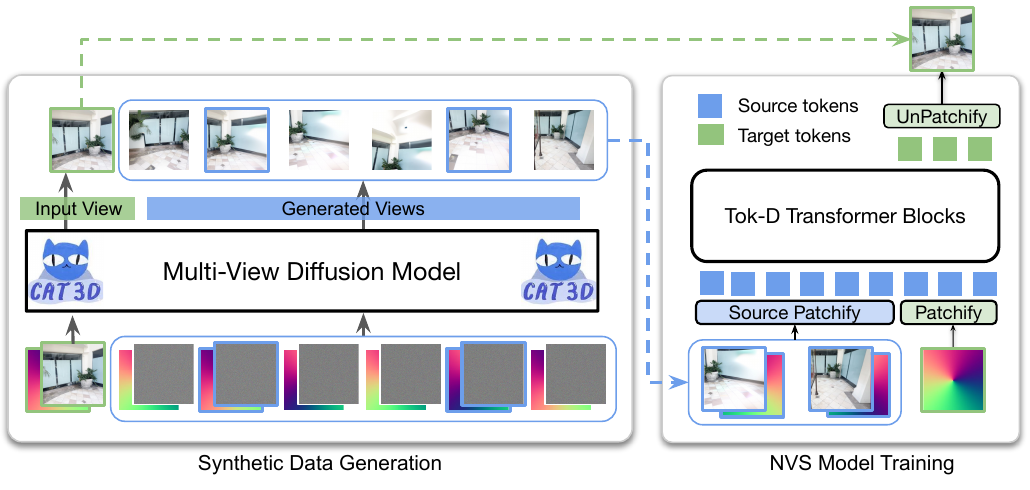}
    \vspace{-5mm}
    \caption{\textbf{An illustration of the architecture.} We use CAT3D, a multi-view diffusion model, to generate synthetic views conditioned on random spline camera trajectories and a random image. From the two random views form the generated views as the source views and the input conditioning view to be the target of our large reconstruction network. Our large reconstruction model uses a special transformer block which we name Tok-D Transformer. When real data is available, we just use the reconstruction transformer.}
    \vspace{-6mm}
    \label{fig:mainfig}
\end{figure*}
\subsection{Offline Novel View Synthesis}
The advent of neural rendering in recent years has substantially improved the quality of NVS. Early neural scene representations focused on the 4D plenoptic function \cite{Gortler1996Lumigraph,Levoy1996} that represents the lightfield of a scene \cite{sitzmann2021lfns,attal2022learning,suhail2022generalizable}. Other methods modeled the geometry of the scene (e.g. as a signed distance function) separately from material properties \cite{wang2021neus, yariv2021volumerenderingneuralimplicit}. Either way, a differentiable rendering process was used to render these neural representations into 2D images \cite{mildenhall2020nerf}. Most of these methods focused on fitting neural fields to sparse observations of a scene at test time---we refer to this as test-time or offline optimization. There is a substantial amount of heterogeneity in these methods, both in terms of the rendering method and the scene representation used. Multi-layer perceptrons (MLPs) \cite{mildenhall2020nerf}, voxels \cite{liu2020neural,yu_and_fridovichkeil2021plenoxels}, hashing-based representations \cite{mueller2022instant, barron2023zipnerf}, triplanes \cite{Chen2022ECCV}, and, most recently, Gaussian splats \cite{kerbl3Dgaussians, Huang_2024, radl2024stopthepop, kheradmand2025stochasticsplatsstochasticrasterizationsortingfree} have been used as scene representations. These methods have trade-offs between reconstruction quality, training time, inference time, memory/space requirements, capacity to model view-dependent effects, etc. Some of these offline methods can even fit dynamic scenes.
These test-time optimization methods demonstrate compelling results given the sparsity of the observations provided. However, they often struggle to incorporate priors learned from larger datasets.

\subsection{Online Novel View Synthesis}
Sometimes referred to as ``feed-forward'' or 
``generalizable'' NVS models, these methods attempt to directly produce 3D representations from input images. Early efforts include the image-based rendering-inspired IBRNet \cite{wang2021ibrnet}, which directly produces 2D images based on epipolar correspondences on the viewing ray. The Large Reconstruction Model (LRM) \cite{hong2024lrm} family of methods attempt to produce a triplane that represents an object, in some cases with near-real time performance. PixelSplat \cite{charatan2024pixelsplat}, MVSplat \cite{charatan2024pixelsplat}, and GS-LRM \cite{Zhang2024GSLRM} attempt to predict 3DGS \cite{kerbl3Dgaussians} representations, which exploit the sparse Gaussian splat representation and fast rasterization to achieve quasi-interactive inference. These methods are trained on large datasets of real-world scenes, which helps them outperform even test-time optimization methods. Quark \cite{Flynn2024} couples an easily-rasterizable layered depth map representation with a render-and-refine strategy to achieve state-of-the-art quality at a much higher resolution.  Other efforts in this space include GPNR \cite{suhail2022generalizable} and SRT \cite{srt22}, which are parameterized in a similar fashion to IBRNet \cite{wang2021ibrnet} and attempt to scale up the image and ray transformers. LRF \cite{kingma2013auto} attempts to perform 3D reconstruction in the latent space of a VAE, bypassing learning 3D representation altogether \cite{zhou2025lrf}. Finally, the LVSM \cite{jin2024lvsmlargeviewsynthesis} removes all 3D priors by simply using one transformer to perform NVS. LVSM performs favorably compared to both geometry-free and geometry-based feed-forward models.
\vspace{-1mm}
\subsection{Synthetic Data}
Recent efforts have leveraged synthetic data to train existing feed-forward NVS methods and investigate its efficacy as a training dataset. However, it is important to note that the synthetic data in many of these efforts are generated procedurally from systems like Blender, whereas ours are generated from a multi-view diffusion model. Two recent works LRM-Zero \cite{xie2024lrmzero} and MegaSynth \cite{jiang2024megasynth} are examples of models trained either entirely or mostly on procedurally generated synthetic data. In LRM-Zero, they demonstrate that the LRM model can be trained entirely on synthetic data. However, the synthetic-data-only model shows a substantial decrease in reconstruction quality compared to the real-world-data equivalent. Improving training data diversity using synthetic data for 4D generation has also been explored in CAT4D \cite{wu2024cat4d}.




 \section{Background}
 \begin{figure*}
    \centering
    \includegraphics[width=\linewidth]{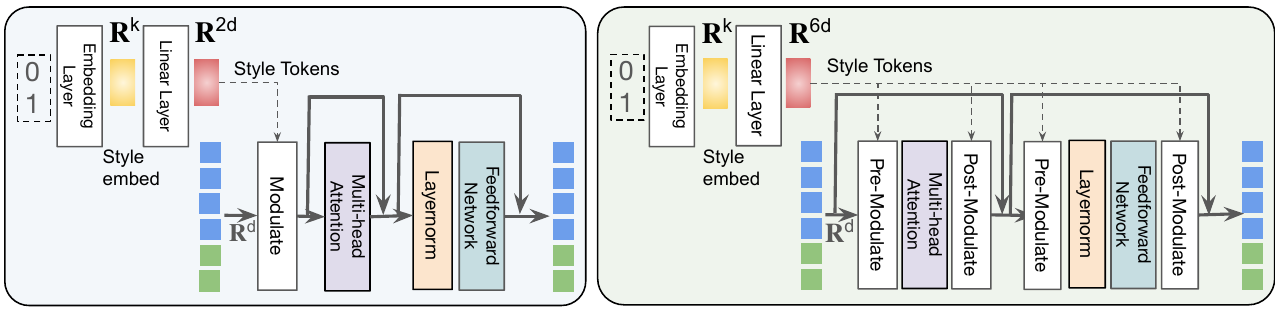}
    \vspace{-8mm}
    \caption{\textbf{An illustration of the Tok-D transformer block.} Our transformer blocks that differentiates between source and target tokens. Tok-D transformer modulates the input to all transformer blocks. Tok-D plus transformer modulates the attention and MLP layers.}
    \vspace{-6mm}
    \label{fig:transformer-block}
\end{figure*}
LVSM is a feed-forward NVS method that has no 3D inductive bias. Since our model builds upon its architecture, we outline the details here for clarity. Where \textsl{i} denotes the image index and \textsl{j} denotes the token index, source images patches are written as $I^s_{ij} \in R^{p\times p \times 3}$, source Plücker coordinates patches $P^s_{ij}\in R^{p\times p \times 6}$, and target Plücker coordinates $P^t_{j}\in R^{p\times p \times 6}$. The source images and plücker embeddings are tokenized together using a linear layer embedder.

\begin{equation}
    S_{ij} =Linear([I^s_{ij}, P^s_{ij}])
\end{equation}
The target Plücker coordinates are also embedded using a linear layer.
\begin{equation}
    T_{ij} =Linear( P^t_{ij})
\end{equation}
Finally, the transformer network is trained to reconstruct the target output tokens  $O^t_{j}$ from the Plücker patch embeddings.
\begin{equation}
    O^t_{j} =M(T_{j}|S_{ij})
\end{equation}
The target output tokens are detokenized using a linear layer which is converted to target image embeddings $T_{j} \in R^{p\times p \times 3} $
\begin{equation}
    T_{j} =Linear([O^t_{j}])
\end{equation}
The target patches are unpatchified to get the target image $T \in R^{H\times W \times 3}$ (see \Cref{fig:mainfig}). The training is supervised using MSE loss and perceptual loss designed to reconstruct.

\noindent \textbf{Transformer Block}
Consider a transformer block at layer $l$, which includes a \emph{Multi-head Self Attention} layer ($\text{SelfAttn}_l$), a Feed-forward network ($\text{FFN}_l$), and a Layer Norm operation ($LN_l$). For an input $[\mathbf{x}^s_l, \mathbf{x}^t_l]$, where $\mathbf{x}^s_l$ and $\mathbf{x}^t_l$ represent the source and target tokens, the data flow as follows:

\begin{align}
[\mathbf{x}^s_l, \mathbf{x}^t_l] &= [\mathbf{x}^s_l, \mathbf{x}^t_l] + \text{SelfAttn}_l([\mathbf{x}^s_l, \mathbf{x}^t_l]) \\ \notag
[\mathbf{x}^s_l, \mathbf{x}^t_l] &= [\mathbf{x}^s_l, \mathbf{x}^t_l] + \text{FFN}_l(LN_l([\mathbf{x}^s_l, \mathbf{x}^t_l])).
\end{align}

Given the basic self attention based transformer blocks in LVSM. At the end of the optimization process there arises a need for all token outputs of a particular layer to be aligned since they are processed by the same set of weights. Hence, LVSM inherently has a chance to infuse noise or atifacts that maybe present in the source images to the target. Moreover this alignment also causes some part of the computational power of the model being used to model source token information although those tokens are discarded at the last layer. Hence, we call for a need to distinguish between the source and target tokens of the transformer network.

\section{Method}

Our proposed method consists of two major contributions. First, our \emph{Token-Disentangled (Tok-D)} transformer block is specialized for NVS and distinguishes information from the source and target views, leading to more efficient allocation of representation capacity. Second, to address the scarcity of multi-view data, we generate synthetic data using CAT3D~\cite{gao2024cat3d} and propose a model training scheme that is robust to artifacts in this synthetic data. In this section, we describe each component in detail.


\subsection{Token-Disentangled Transformer}

In LVSM, the transformer blocks process source and target tokens in the same manner, even though the source consists of images and Plücker rays, while the target includes only Plücker rays. Additionally, source and target image quality can differ when training with synthetic data. To address this, we introduce the \textit{Token-Disentangled (Tok-D) Transformer} block (see \Cref{fig:transformer-block}), which enables differentiated processing of source and target tokens through modulation. The Tok-D Transformer uses an indicator variable ($\delta$), where $\delta = 1$ for target tokens and $\delta = 0$ for source tokens, to modulate tokens based on their origin. This mechanism extracts distinct style vectors and computes specific scale and bias parameters for each layer and token type, allowing for precise and adaptive token modulation.
\begin{align}
\mathbf{style} &= \text{Linear}(\text{Embed}(\delta)) \\ \notag
\text{Mod}_l(\mathbf{x}) &= (1+\mathbf{\sigma}_l)\mathbf{x} + \mathbf{\mu}_l, \text{where\ } [\mathbf{\sigma}_l, \mathbf{\mu}_l] = \text{Linear}_l(\mathbf{style}) \\ \notag
[\mathbf{x}^s_l, \mathbf{x}_l^t] &= \text{Mod}^{s,t}_l([\mathbf{x}^s_l, \mathbf{x}_l^t]) = [\text{Mod}^s_l(\mathbf{x}^s_l),\text{Mod}^t_l(\mathbf{x}^t_l)]
\end{align}
 Modulating the input of each transformer block improves perforamnce. Drawing inspiration from DiT \cite{Peebles2022DiT}, we extend this modulation to the Attention and MLP layers, achieving further improvements. This modulation is termed \textit{pre-modulation} if applied before a layer and \textit{post-modulation} if after. Pre-modulation includes both scaling and shifting, and post-modulation involves only scaling.
\begin{align}
[\hat{\mathbf{x}}^s_l, \hat{\mathbf{x}}_l^t] &= \text{Mod}^{s,t}_{l1}([\mathbf{x}^s_l, \mathbf{x}_l^t]) \\ \notag
[\mathbf{x}^s_l, \mathbf{x}_l^t] &= [\mathbf{x}^s_l, \mathbf{x}_l^t] + [\sigma^{s}_{l1}, \sigma^{t}_{l1}] \odot \text{SelfAttn}([\hat{\mathbf{x}}^s_l, \hat{\mathbf{x}}_l^t]) \\ \notag
[\hat{\mathbf{x}}^s_l, \hat{\mathbf{x}}_l^t] &= \text{Mod}^{s,t}_{l2}([\mathbf{x}^s_l, \mathbf{x}_l^t]) \\ \notag
[\mathbf{x}^s_l, \mathbf{x}_l^t] &= [\mathbf{x}^s_l, \mathbf{x}_l^t] + [\sigma^{s}_{l2}, \sigma^{t}_{l2}] \odot \text{FFN}_l(\text{LN}_l([\hat{\mathbf{x}}^s_l, \hat{\mathbf{x}}_l^t]))
\end{align}
where $\odot$ denotes element-wise multiplication which scales the corresponding source and target tokens.

Our Tok-D transformer block enhances the distinction between source and target tokens, as reflected in their distinct feature representations (\Cref{fig:sidebyside}, Section~\ref{sec:analysis}). This specialization highlights the superior representational capacity of our model. Furthermore, when trained on synthetic data (Section~\ref{sec:syntheticpipline}), out-of-distribution artifacts can introduce quality disparities between source and target tokens. By leveraging its token-aware architecture, our model demonstrates greater robustness to these artifacts, resulting in improved performance, as shown in Section~\ref{sec:cross_dataset}.

\subsection{Synthetic Data Generation \& Training Scheme}
\label{sec:syntheticpipline}
\begin{figure*}
    \centering
    \includegraphics[width=\linewidth]{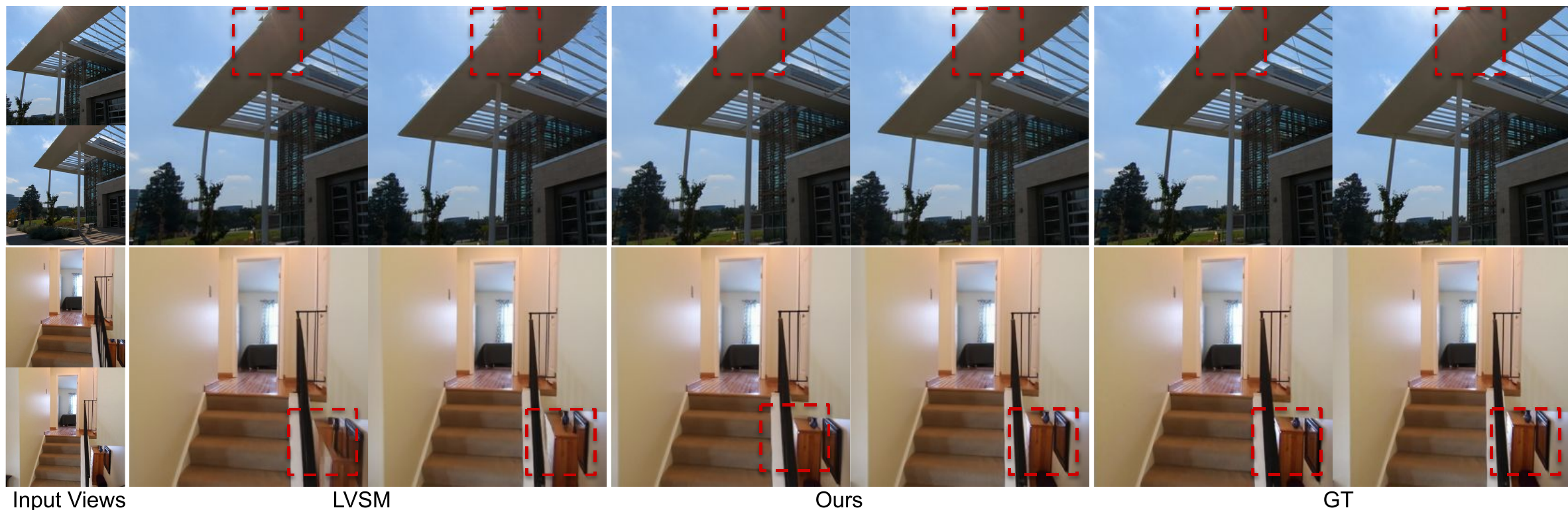}
    \vspace{-6mm}
    \caption{\textbf{Qualitative results on in-distribution datasets.} We illustrate the cases Tok-D transformer works better than LVSM. We notice that we obtain substantial improvement in cases where the novel views needs to reconstruct regions present only in one of the views as shown in the highlighted regions in the images. The results presented here are taken from our in-distribution trained model. We present two diffrent views to show that this problem is persistent across views.}
    \label{fig:indistcompate}
    \vspace{-3mm}
\end{figure*}

Training a naive transformer model with synthetic data can lead to degraded performance rather than improvement due to two key factors: (1) The model struggles to distinguish between tokens from source images and target images, allowing artifacts from one to propagate into the other during alignment. (2) The model is trained to generate novel views from sparse input views, and if the target is a synthetic image with artifacts, it may learn a distribution biased toward unrealistic images. While these issues might not arise with perfect synthetic data, in-practice synthetic datasets often contain noise, making the model vulnerable to errors through either mechanism.
However, for image-to-multiview synthesis models like CAT3D, we propose a simple yet effective solution: assigning the conditioned image as the target view and the generated views as input views.

Formally let $I_c,C_c$ denote the input image and camera conditioning used for the multiview diffusion model. We sample  additional random spline camera trajectory poses $C_{tgt}$ relative to this particular view, and use the state-of-the-art multi-view diffusion model CAT3D to generate the target views $I_{src}$ conditioned on the input conditioning and target poses
\begin{equation}
I_{gen} \sim DM(I_{gen}|C_{gen},C_c,I_c)    
\end{equation}
Here DM represents inferencing through the state of the art diffusion model, After obtaining the generated views, we sample 2 generated views $I_{src}$
\begin{equation}
I_{src},C_{src} \sim I_{gen},C_{gen} 
\end{equation}

and their camera poses as the source images $I_{src}, C_{src}$ and utilize the conditioned image and its camera as the target $I_c,C_c$. Sampling the source and target images this way forces the transformer to always generate a realistic image, making our model robust to artifacts from synthetic data. 


\definecolor{lightyellow}{rgb}{1,1, 0.8}
\definecolor{yellow}{rgb}{1,0.97, 0.65}
\definecolor{orange}{rgb}{1, 0.85, 0.7}
\definecolor{tablered}{rgb}{1, 0.7, 0.7}

\begin{table*}[t!]
\caption{\textbf{Quantitative comparisons for in-distribution scene synthesis at 256 resolution.}  LVSM and our method are trained with a batch size of 64. LVSM results are taken from the original paper rather than our re-implementation. Our method outperforms the previous SOTA method across all exisiting datasets. (\textcolor{tabfirst}{\rule{10pt}{10pt}},\textcolor{tabsecond}{\rule{10pt}{10pt}}, \textcolor{tabthird}{\rule{10pt}{10pt}})  denotes the first, second and third best results.} 
\vspace{-0.1in}
\resizebox{0.99\linewidth}{!}{
\begin{tabular}{r|c|ccc|ccc|ccc} 
    Method& Venue &  \multicolumn{3}{c}{RealEstate10k~\citep{Zhou2018}}&  \multicolumn{3}{c}{ACID~\citep{infinite_nature_2020}}&  \multicolumn{3}{c}{DL3DV~\citep{ling2024dl3dv}}  \\ 
    && PSNR \(\uparrow\) & SSIM \(\uparrow\)  & LPIPS \(\downarrow\) & PSNR \(\uparrow\) & SSIM \(\uparrow\)  & LPIPS \(\downarrow\) & PSNR \(\uparrow\) & SSIM \(\uparrow\)  & LPIPS \(\downarrow\)  \\ 
    \midrule
    GPNR~\citep{suhail2022generalizable} &CVPR'23&24.11& 0.793 &0.255& 25.28 &0.764 &0.332&-&-&-\\
    PixelSplat~\citep{charatan2024pixelsplat} &CVPR'24& 25.89 & 0.858  & 0.142 & 28.14 & 0.839  & 0.533 & - & -  & -\\
 MVSplat~\citep{chen2024mvsplat} & ECCV'25&26.39 & 0.869  & 0.128& \cellcolor{tabthird}28.25 &\cellcolor{tabthird} 0.843  & \cellcolor{tabthird}0.144 & 17.54 & 0.529  & 0.402  \\
    DepthSplat ~\citep{xu2024depthsplat} &CVPR'25&\cellcolor{tabthird}27.44& \cellcolor{tabthird}0.887  &\cellcolor{tabthird} 0.119 & -& - & -& \cellcolor{tabthird}19.05& \cellcolor{tabthird}0.610  &\cellcolor{tabthird} 0.313\\
     LVSM ~\citep{jin2024lvsmlargeviewsynthesis}& ICLR'25&\cellcolor{tabsecond}28.89 &\cellcolor{tabsecond} 0.894 &\cellcolor{tabsecond} 0.108&\cellcolor{tabsecond}29.19 & \cellcolor{tabsecond}0.836  & \cellcolor{tabsecond}0.095 &\cellcolor{tabsecond} 19.91 & \cellcolor{tabsecond}0.600 & \cellcolor{tabsecond}0.273  \\
     Ours& & \cellcolor{tabfirst}30.02 & \cellcolor{tabfirst}0.919 &\cellcolor{tabfirst}0.058& \cellcolor{tabfirst}29.47 &\cellcolor{tabfirst}0.846 & \cellcolor{tabfirst}0.086 &\cellcolor{tabfirst}21.55& \cellcolor{tabfirst}0.643
     & \cellcolor{tabfirst}0.208  \\
     \bottomrule
\end{tabular}
}
\label{tab:compare_scenes}
\vspace{-0.2in}
\end{table*}

\section{Experiments}

\subsection{Implementation Details}

\textbf{Training details} We perform all experiments on 8 H100 GPUs. We use the AdamW optimizer with $\beta$ parameters $0.9$ and $0.95$, and we use weight decay with a rate of $0.05$ for all layers except the normalization layers. Moreover, we use a linear learning rate scheduler with with a peak learning rate of $2e^{-4}$, and a warmup of 2500 iterations. In total, all experiments have $100k$ training iterations. In addition, we use exponential moving averaging (EMA) with a rate of $0.99$ for stabilizing the training process. Although previous works required gradient clipping for a stable training process, our training processes were smooth without a need for an explicit gradient clipping.

\noindent\textbf{Training and Evaluation Datasets} 
    For scene-level synthesis model training, we use Re10K \cite{Zhou2018}, ACID \cite{infinite_nature_2020} and DL3DV \cite{ling2024dl3dv} with their originally released train and test splits. We also perform an experiment where the model is trained together with a mix of all of these datasets. For scene-level synthesis, we follow LVSM and train using 2 input views and test using 6 target views fed one at a time. For DL3DV dataset evaluation, we choose the farthest camera from a randomly selected target view as the input view. The training and evaluation of DL3DV dataset for in distribution metrics is done using 2 input views and 2 target views. For cross dataset testing, we use 2 input views and 6 target views for DL3DV dataset. We use a batch size of 64 for our experiments.

\noindent\textbf{Synthetic Data} For generating the synthetic data we use the state-of-the-art 3D generation model CAT3D. CAT3D was trained using a single scene dataset Re10K and three object-based datasets: Objaverse \cite{objaverse}, MVImgNet \cite{yu2023mvimgnet} and Co3D \cite{reizenstein2021common}. To create synthetic data, we use two variants: one with 1 conditioning view and 7 generated views, and another with 3 conditioning views and 5 generated views. We match the focal lengths of Re10K and DL3DV during generation. For the camera trajectory, we sample a random spline trajectory with a random position rotation matrix, converting it into ray maps before passing it into the network. As CAT3D is originally trained with a resolution of 512, we convert the images and camera parameters to a resolution of $256$ before passing them through our network.

\subsection{Scene Synthesis}
\vspace{-2mm}
\begin{figure*}[t!]
    \centering
    \includegraphics[width=\linewidth]{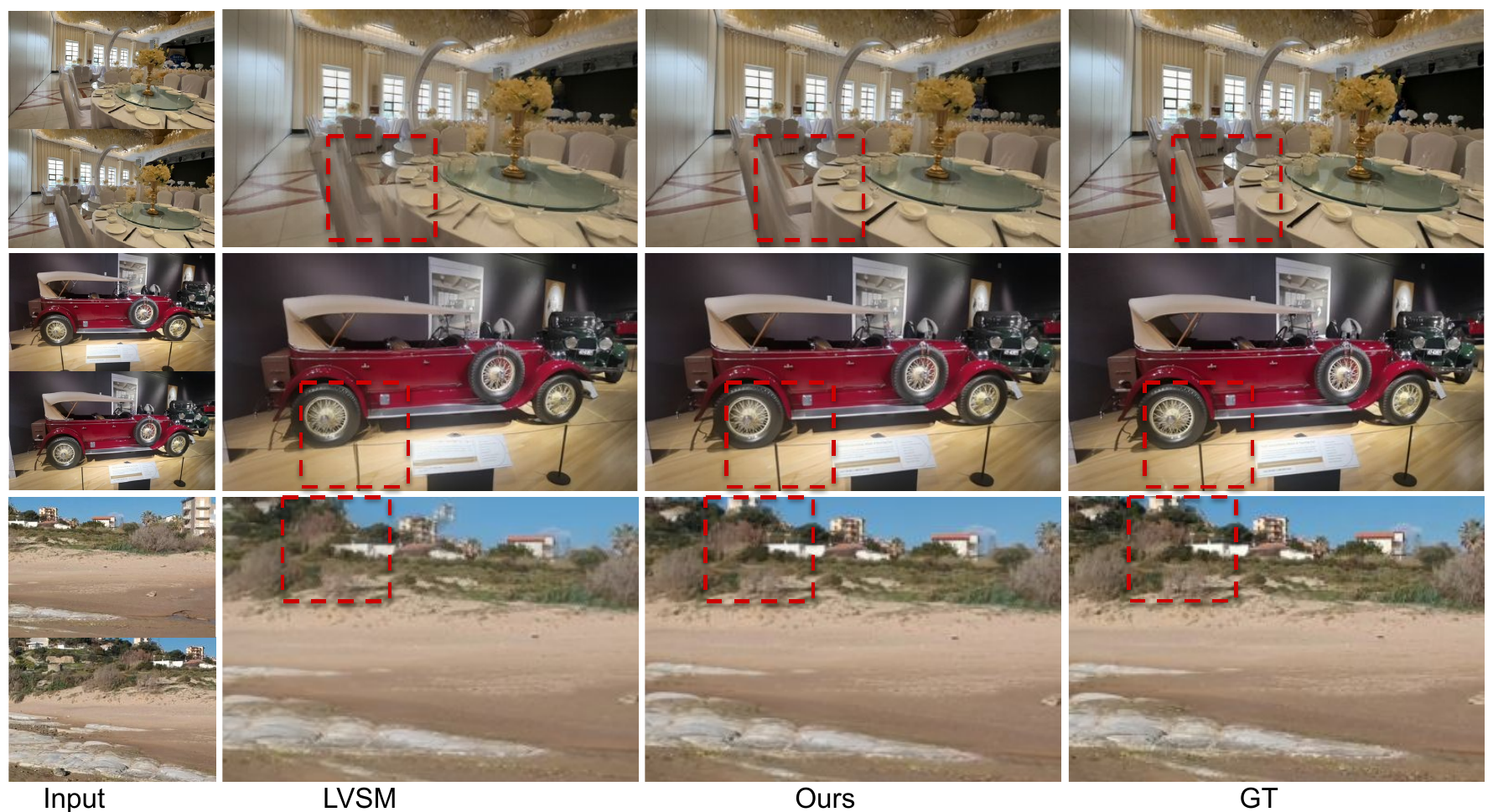}
    \vspace{-6mm}
    \caption{\textbf{Out-of-distribution Evaluation:} We evaluate our the version of our method fine-tuned on synthetic data and LVSM on DL3DV and ACID (i.e. out-of-distribution datasets). We also evaluate the model with resolutions that were not used during training. We notice that LVSM's visual quality degrades when substantial camera motion reveals previously-occluded regions.}
    \label{fig:my_label}
\end{figure*}
We evaluate our method qualitatively and quantitatively for scene synthesis using very recent feed-forward methods  GPNR, PixelSplat, MVSplat, DepthSplat and LVSM. These methods were chosen because they outperform conventional approaches in 2-view reconstruction. Quantitative results are shown in \Cref{tab:compare_scenes}. We observe that Tok-D-Plus outperforms LVSM by 1.2 dB on the Re10K evaluation benchmark when both models are trained with 8 GPUs. Furthermore, despite using only 8 GPUs, our method still surpasses LVSM trained with 64 GPUs by a margin of 0.2 dB. Moreover we obtain an improvement of 1.6dB over LVSM in a more diverse scene-level dataset,  DL3DV~\citep{ling2024dl3dv} dataset as well. We also observe that our performance improvement is $0.2$ in ACID dataset. We emphasize that this happens because ACID has a relatively smaller training and testing set and the dataset is generally clean and easier to reconstruct. We also provide the corresponding qualitative comparisons on Re10K and DL3DV dataset in \Cref{fig:indistcompate} . Comparing the main results we find that our method usually outperforms LVSM when the generated content is only visible in one of the source views. When the camera is far from both views and the information is present only in one of the views, our method is still able to extract the relevant content from the corresponding input image. As can be seen from rows 1 and 2, the reconstruction form LVSM fails to reconstruct objects present in only one of the views, whereas Tok-D transformer can effectively reconstruct these regions. 

\definecolor{lightyellow}{rgb}{1,1, 0.8}
\definecolor{yellow}{rgb}{1,0.97, 0.65}
\definecolor{orange}{rgb}{1, 0.85, 0.7}
\definecolor{tablered}{rgb}{1, 0.7, 0.7}

\begin{table*}[t!]
\vspace{-4mm}
\caption{\textbf{Quantitative comparisons for scaling up with synthetic data.} We evaluate LVSM and our method, which are both trained with a batch size of 64. A mixture of synthesized DL3DV and Re10K data is used for the synthetic tab. For MVSplat and DepthSplat we include the numbers reported in their papers}
\vspace{-0.1in}
    

\resizebox{0.99\linewidth}{!}{
\begin{tabular}{r|cc|ccc|ccc|ccc} 
    Method&\multicolumn{2}{c}{Training Dataset}&  \multicolumn{3}{c}{RealEstate10k~\citep{Zhou2018}}&  \multicolumn{3}{c}{ACID~\citep{infinite_nature_2020}}&  \multicolumn{3}{c}{DL3DV~\citep{ling2024dl3dv}}  \\ 
     & Re10K\cite{Zhou2018}&Synthetic & PSNR \(\uparrow\) & SSIM \(\uparrow\)  & LPIPS \(\downarrow\) & PSNR \(\uparrow\) & SSIM \(\uparrow\)  & LPIPS \(\downarrow\) & PSNR \(\uparrow\) & SSIM \(\uparrow\)  & LPIPS \(\downarrow\)  \\ 
    \midrule 
     MVSplat~\citep{chen2024mvsplat} &\checkmark &&26.39&0.869&0.128& 28.15 &
0.147&
0.841 &
17.72&0.534&0.371\\
     Depthsplat~\citep{xu2024depthsplat} &\checkmark& &27.44&0.887&0.119&-&-&-&18.90&0.640&0.317\\
     
     LVSM~\citep{jin2024lvsmlargeviewsynthesis}& \checkmark &&\cellcolor{tabthird}28.89 & \cellcolor{tabthird}0.894 & \cellcolor{tabthird}0.108 &\cellcolor{tabthird} 28.29 & \cellcolor{tabthird}0.809  &\cellcolor{tabthird}0.104 & \cellcolor{tabthird}20.52 &\cellcolor{tabthird} 0.621  &\cellcolor{tabthird} 0.223  \\
     LVSM~\citep{jin2024lvsmlargeviewsynthesis} & \checkmark  &\checkmark&    28.49 & 0.892  & 0.070 &  28.16 &0.802  &0.107  &  20.21 & 0.595  & 0.240   \\
     Ours & \checkmark &  &\cellcolor{tabfirst} 30.02 & \cellcolor{tabfirst}0.910  & \cellcolor{tabfirst}0.058 &    \cellcolor{tabsecond}29.31&
 \cellcolor{tabsecond}0.838&
 \cellcolor{tabsecond}0.091
  &   \cellcolor{tabsecond}21.18 &  \cellcolor{tabsecond}0.652  &  \cellcolor{tabsecond}0.205   \\
     Ours & \checkmark &\checkmark   &   \cellcolor{tabsecond}29.97&
\cellcolor{tabsecond}0.920&
\cellcolor{tabsecond}0.058
  &  \cellcolor{tabfirst}29.37& \cellcolor{tabfirst}0.839 &\cellcolor{tabfirst}0.091 & \cellcolor{tabfirst}21.27&\cellcolor{tabfirst}0.657&\cellcolor{tabfirst}0.202   \\

\bottomrule
\end{tabular}
}
\label{tab:compare_scenescross}
\vspace{-0.27in}
\end{table*}

\subsection{Cross-Dataset Scene Synthesis}\label{sec:cross_dataset}
To analyze the generalization capacity of our method, we evaluate our method trained with Re10K dataset on two different datasets: ACID and DL3DV~\citep{ling2024dl3dv}. ACID is a dataset with aerial views similar to Re10K. DL3DV~\citep{ling2024dl3dv} is a diverse dataset comprising natural scenes and various indoor and outdoor settings. The scene geometry and appearance of DL3DV~\citep{ling2024dl3dv} is very different from Re10k. We test the  Re10K and ACID datasets at a resolution of 256$\times$265. For testing on DL3DV~\citep{ling2024dl3dv}, we choose a resolution of 256$\times$448 to maintain the original aspect ratio in the DL3DV~\citep{ling2024dl3dv} dataset and well as maintain consistent evaluation settings with DepthSplat. We choose 2 source views and 6 target views for all of these datasets. Looking closely at the quantitative results on \Cref{tab:compare_scenes} and \Cref{tab:compare_scenescross}, we find that the model trained on Re10K underperformed the in-distribution trained model by a small margin. The drop is higher in the case of DL3DV due to resolution and diversity differences in the datasets. Next we add a small portion of synthetic data comprising about half the size of Re10K dataset and perform training with the new framework. We also retrain LVSM for the same setting. We find that \textit{LVSM's performance drops rather than improving when synthetic data is added}. We emphasize that this arises due to the introduction of artifacts during feature alignment. In contrast, we observe an improvement in quality on our method when a small amount of synthetic data is added.

\begin{figure*}[htbp]
    \centering
\begin{minipage}[t]{0.125\linewidth}
    \centering
    \begin{subfigure}[c]{0.49\linewidth}
        \centering
        \includegraphics[width=\linewidth]{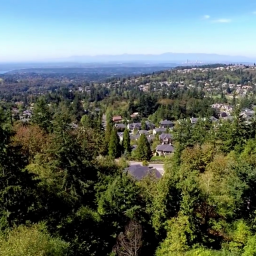}
        \subcaption{Target}
        \label{fig:tgt_images}
    \end{subfigure}\hfill
    \begin{subfigure}[c]{0.49\linewidth}
        \centering
        \includegraphics[width=\linewidth]{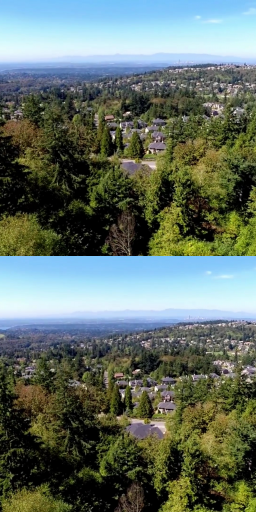}
        \subcaption{Source}
        \label{fig:src_images}
    \end{subfigure}\hfill
\end{minipage}\hfill
    \begin{minipage}[t]{0.87\linewidth}
        \centering
        \begin{subfigure}[c]{0.245\linewidth}
            \centering
            \includegraphics[width= \linewidth]{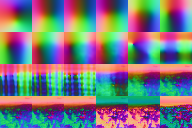}
            \subcaption{LVSM Target PCA}
            \label{fig:tgt_lvsm_pca}
        \end{subfigure}\hfill
        \begin{subfigure}[c]{0.245\linewidth}
            \centering
            \includegraphics[width=\linewidth]{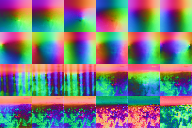}
            \subcaption{LVSM Source PCA}
            \label{fig:src_lvsm_pca}
        \end{subfigure}\hfill
        \begin{subfigure}[c]{0.245\linewidth}
            \centering
            \includegraphics[width=\linewidth]{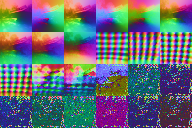}
            \subcaption{Ours Source PCA}
            \label{fig:src_lvsm_pca}
        \end{subfigure}\hfill
        \begin{subfigure}[c]{0.245\linewidth}
            \centering
            \includegraphics[width= \linewidth]{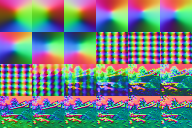}
            \subcaption{Ours Target PCA}
            \label{fig:tgt_lvsm_pca}
        \end{subfigure}\hfill
    \end{minipage}
    \vspace{-3mm}
    \caption{\textbf{A visualization of the principal components of transformer layer outputs for source and target of LVSM}. The 24 images in each subfigure show the layer output of each layer of the transformer. LVSM features for source and target images looks similar even though the source is conditioned with image and Plücker coordinates and target is conditioned with Plücker coordinates alone. This leads to inefficient transformer usage requiring explicit alignment of source and target features across different layers  }
    \label{fig:sidebyside}
\end{figure*}

 \definecolor{lightyellow}{rgb}{1,1, 0.8}
\definecolor{yellow}{rgb}{1,0.97, 0.65}
\definecolor{orange}{rgb}{1, 0.85, 0.7}
\definecolor{tablered}{rgb}{1, 0.7, 0.7}

\begin{table*}[t!]
\vspace{-4mm}
\caption{\textbf{Ablation studies on scaling up with more real data.
}Although including synthetic data in training is helpful for improving quality, including additional real data significantly improves reconstruction quality.}
\vspace{-0.1in}
    

\resizebox{0.99\linewidth}{!}{
\begin{tabular}{r|cc|ccc|ccc|ccc} 
    Method&\multicolumn{2}{c}{Training Dataset}&  \multicolumn{3}{c}{RealEstate10k~\citep{Zhou2018}}&  \multicolumn{3}{c}{ACID~\citep{infinite_nature_2020}}&  \multicolumn{3}{c}{DL3DV~\citep{ling2024dl3dv}}  \\ 
     & Re10K~\citep{Zhou2018}$+$ Synthetic&DL3DV~\citep{ling2024dl3dv} & PSNR \(\uparrow\) & SSIM \(\uparrow\)  & LPIPS \(\downarrow\) & PSNR \(\uparrow\) & SSIM \(\uparrow\)  & LPIPS \(\downarrow\) & PSNR \(\uparrow\) & SSIM \(\uparrow\)  & LPIPS \(\downarrow\)  \\ 
    \midrule 
     LVSM~\citep{jin2024lvsmlargeviewsynthesis} & \checkmark & &   28.49 & 0.892  & 0.070 &  28.16 &0.802  &0.107  &  20.21 & 0.595  & 0.240   \\
LVSM~\citep{jin2024lvsmlargeviewsynthesis} & \checkmark &\checkmark  &  28.10 &0.892&
0.073
  &  28.79& 0.826 &0.096& 21.37&
0.665&0.196\\
   
     Ours & \checkmark  &  &   \cellcolor{tabfirst}29.97&
\cellcolor{tabfirst}0.920&
\cellcolor{tabfirst}0.058
  &  \cellcolor{tabsecond}29.37& \cellcolor{tabsecond}0.839 &\cellcolor{tabsecond}0.091 & \cellcolor{tabsecond}21.27&\cellcolor{tabsecond}0.657&\cellcolor{tabsecond}0.202   \\
    Ours & \checkmark  &\checkmark &   \cellcolor{tabsecond}29.78&
\cellcolor{tabsecond}0.917&
\cellcolor{tabsecond}0.0604
  &  \cellcolor{tabfirst}30.13& \cellcolor{tabfirst}0.857 &\cellcolor{tabfirst}0.082 & \cellcolor{tabfirst}23.14&\cellcolor{tabfirst}0.726&\cellcolor{tabfirst}0.156   \\
\bottomrule
\end{tabular}
}
\label{tab:compare_realsynthetic}
\vspace{-0.2in}
\end{table*}

 \definecolor{lightyellow}{rgb}{1,1, 0.8}
\definecolor{yellow}{rgb}{1,0.97, 0.65}
\definecolor{orange}{rgb}{1, 0.85, 0.7}
\definecolor{tablered}{rgb}{1, 0.7, 0.7}

\begin{table}[t!]
\centering
\caption{\textbf{Ablation analysis} We analyze the performance improvement of our design choices. \textit{Pre} and \textit{Post} demonstrate the effects of including or not including pre/post-modulation.}
\vspace{-0.1in}
\resizebox{\linewidth}{!}{
\begin{tabular}{cc|ccc|ccc} 
    Pre &Post&Whole&Attn&MLP & PSNR \(\uparrow\) & SSIM \(\uparrow\)  & LPIPS \(\downarrow\)   \\ 
    \midrule 
    \midrule 
     &&&& &28.50 &0.893&0.070   \\
     \checkmark &&\checkmark&&& 29.69&0.911 &  0.063   \\
     \checkmark &&&\checkmark& \checkmark& 28.51& 0.894 &0.070   \\

     \checkmark &\checkmark&&\checkmark&\checkmark & 30.02 & 0.918  & 0.058    \\
     \bottomrule
\end{tabular}
}
\label{tab:compare_ablation}
\vspace{-0.2in}
\end{table}


\vspace{-2mm}

\subsection{Analysis and Discussion}
\label{sec:analysis}
\noindent\textbf{Visualization of source and target features.} To visually illustrate the representation alignment problem mentioned in the previous sections, we visualize the 3 channel PCA of each transformer block output after unpatchifying for all 24 layers of LVSM and our method in \Cref{fig:sidebyside}. The first row shows the first 6 layer outputs, second row shows layer 6-12, and so on. We can see that for a particular scene the source and target layer tokens are aligned at all layers even though the training objective is to reconstruct the target. This causes inefficient usage of the transformer parameters to maintain the source information throughout the layers. Moreover this also makes the model prone to \textit{noise in the source data}. However, with our Tok-D transformer there is no alignment and the source information is infused much earlier, leaving more room for the transformer blocks to reconstruct the target. Another important observation is that although both source image and Plücker coordinates are fed as input to the source, the source tokens look similar to the Plücker coordinates. Whereas in our case the image components in the source PCA components leading to much more effective information extraction from each source token.

\noindent\textbf{Impact of additional real data.}  Incorporating synthetic data into the training process facilitates the introduction of diverse scenes and camera motion, enhancing model generalizability. While the proposed Tok-D transformer demonstrates reduced sensitivity to synthetic data artifacts and increased generative diversity, its photorealistic reconstruction performance remains comparable to the baseline model trained solely on real data. To investigate the impact of augmenting the training dataset with additional real data, we integrated the DL3DV dataset into the existing experimental setup. This modification resulted in a significant improvement in photorealistic reconstruction, as evidenced by a substantial increase in PSNR on the ACID dataset. Furthermore, the relative performance gains observed with our model, compared to LVSM, were considerably greater, suggesting a reduced susceptibility to noise.

\vspace{-1mm}

\subsection{Ablation Studies}
\vspace{-1mm}
We analyze the impact of various design choices in the network. Specifically, we examine three aspects: (1) The effect of modulation in different parts of the network, (2) The role of EMA in performance, (3) Number of input views.

\noindent\textbf{Impact of modulation at different locations  of Tok-D transformer.} We examine the effect of modulating different parts of the network. For this, we consider four different cases. We present the corresponding results in \Cref{tab:compare_ablation}. Having a common modulation premodulation worked better than separate premodulation for both layers.

\noindent\textbf{Impact of EMA.} We also observe that performing Exponential moving average (EMA) \cite{haynes2012exponential}  during training results in a performance boost for the base model. For the sake of consistency, we show the results of our model and our re-implementation of LVSM with 1024 channels trained with and without EMA in \Cref{tab:lvsmema}.
\begin{table}[t]
\centering
\caption{\textbf{Effect of EMA on runtime performance and quality.} Comparison performed on Re10k.}
\vspace{-2mm}
\resizebox{0.5\textwidth}{!}{%
\begin{tabular}{c|c|c|c|ccc|ccc}
\toprule
Method & Train & Render & GFLOPs & \multicolumn{3}{c|}{No EMA} & \multicolumn{3}{c}{With EMA} \\
& (ms) & (ms) & & PSNR & SSIM & LPIPS & PSNR & SSIM & LPIPS \\
\midrule
LVSM-1024 & 706.1 & 171.6 & 2896.88 & 27.68 & 0.88 & 0.077 & 28.65 & 0.90 & 0.070 \\
Ours      & 734.6 & 174.4 & 2900.78 & 28.75 & 0.90 & 0.064 & 30.02 & 0.92 & 0.058 \\
\bottomrule
\end{tabular}
}
\label{tab:lvsmema}
\vspace{-3mm}
\end{table}
\begin{table}[t]
\centering
\caption{\textbf{Effect of adding more source views}. Our method works well as additional source views are introduced.}
\vspace{-3mm}
\resizebox{\linewidth}{!}{
\begin{tabular}{c|c c c | c c c| c c c}
\toprule
Method&\multicolumn{3}{c}{2 views} & \multicolumn{3}{c}{4 views}& \multicolumn{3}{c}{8 views}\\
\midrule
& PSNR & SSIM & LPIPS & PSNR & SSIM & LPIPS & PSNR & SSIM & LPIPS \\
\midrule
Ours &  30.02 & 0.92 &0.058&  31.51 & 0.94 & 0.048 & 33.09 & 0.94 & 0.042 \\
\bottomrule
\end{tabular}
}
\label{tab:numviews}
\vspace{-6mm}
\end{table}

\noindent\textbf{Impact of number of source frames.} Our model scales with the number of input views and results in better reconstruction quality when more input views are fed to the model to the model as presented in \Cref{tab:numviews}.
\vspace{-3mm}
\section{Conclusion}
\vspace{-1mm}
In this paper, we introduce a new approach to scaling up NVS by addressing two key limitations in existing models: efficiency and diversity. To enhance the efficiency of transformer-based NVS models, we propose the Token-Disentangled (Tok-D) Transformer, which reduces redundancies and improves data efficiency, enabling higher reconstruction quality with less compute. Additionally, the Tok-D Transformer mitigates training artifacts through its disentangling property, allowing for effective scaling using synthetic data. Incorporating synthetic data into training significantly improves cross-dataset performance compared to existing models. By integrating the Tok-D Transformer and synthetic data, we achieve state-of-the-art results across three large-scale NVS benchmarks, surpassing previous methods with lower computational cost and by a substantial margin.

{
    \small
    \bibliographystyle{ieeenat_fullname}
    \bibliography{main}
}

\section{Design choices}

We provide further details of the exact transformer model used here.
\noindent\textbf{Transformer blocks} We find the claims regarding the naive transformer architecture to be unstable for image generative tasks to be true. We use QK-Norm to stabilize the transformer block. We use $24$ transformer blocks with an embedding dimension of $1024$. In addition to this, different from LVSM, we use attention biases at all layers and include the bias for the last transformer block, as we find this design choice particularly stable with linear learning rate decay. We use a patch size of 8 for all experiments.

\subsection{Enhancing 3D generative models for 3D consistent generation}

The use of diffusion models has been widely explored for generating 3D scenes. Multiple works in the literature adapt pretrained text-to-image and image-to-video models for 3D-consistent scene generation. Most of these works condition the diffusion model on camera parameters and learn the conditional distribution of multiple views given the camera poses. Given the ability to cherry-pick and sample through the diffusion model multiple times, these models produce high-quality results. However, existing 3D scene generation models cannot mass-produce synthetic data for fine-tuning substream models for high-fidelity generation. Until now, no generalizable models with high-fidelity results have been proposed that can directly utilize the data generated by diffusion models. We argue that this drawback is caused by a lack of analysis of the inference-time generation process of diffusion models. Although extensive studies have been performed on different training strategies for 3D-consistent generation using diffusion models, much less effort has been put into improving inference-time generation quality.

Most 3D generative models generate $N$ views of a scene, each of dimension $(H \times W \times C)$, in parallel to preserve 3D consistency. The generation process starts with random isotropic Gaussian noise of dimension $N \times H \times W \times C$, which undergoes a diffusion process of $T$ steps. This either converts it into a latent representation, which is then decoded by a VAE decoder to produce multiview images, or generates images directly. These multiview images are further used to train a NeRF or a Gaussian Splat model to generate novel views of the scene. When the diffusion model generates high-quality, 3D-consistent images, this framework works perfectly. However, in reality, diffusion models are sensitive to input noise. Even for the simple case of image generation, different noise inputs produce different quality results. Recent works have shed light on inference-time scaling laws for generation, claiming that the quality of diffusion model outputs can be controlled by selecting the correct input noise via rejection sampling. Similar claims have been made for video generation models, where performance improves significantly by refining the input noise schedule.

To understand this, consider a toy example: Suppose we want to generate an image ($I_1$) using the diffusion model conditioned on a text prompt. Starting with Gaussian noise $N_1$, if we want to generate another image ($I_2$) close to ($I_1$), the desired noise is most likely closer to $N_1$. Previous works have demonstrated enhanced video generation results by selecting starting noises that are close across different frames.

In our case, we use the image-to-multiview variant of CAT3D as the base model for generating multiview images. For choosing the initial noise, we follow a specific heuristic. Specifically, we ensure that the noise across different views remains 3D-consistent. CAT3D is a multiview diffusion model that generates eight views simultaneously, conditioned on the camera poses. CAT3D allows conditioning on a particular view to generate the remaining views. Given the view to be conditioned, we select a random noise for this view, denoted as $V_1$, with its noise represented as $N_1$ and the corresponding rotation-translation matrices denoted as $R_1, t_1$. To estimate the starting noise for other views, we perform a warping operation on $N_1$, denoted by:

\begin{equation} N_i = warp(N_1, inv([R_i,t_i])[R_1,t_1]) \end{equation}

where the warp operation transforms the coordinates of $N_1$ to $N_i$. However, we noticed that such a warping operation fails in regions outside the scene. To handle these cases while enhancing consistency, we marginally modify the noise. Specifically, for these cases, we assign the noise as:

\begin{equation} N_2 = \alpha N_1 + (1-\alpha) \mathcal{N}(0,I) \end{equation}

Thus, our effective starting noise is defined as:
$$
N_{final}=
\begin{cases}
N_1, \text{overlapping regions}\\
N_2, \text{non overlapping regions}
\end{cases}
$$

We perform the effective noising operation  parallely with respect to the reference view. First we take view 1, warp to view 2. then add noise, then we Although we use CAT3D, our method is generalizable across any 3D scene generation model.

Understanding the value that synthetic data from generative models can bring, we propose a method to enhance diffusion-based 3D generative models to produce high-quality, 3D-consistent results.

\subsection{Loss functions}

Similar to LVSM, we utilize Mean Square Error (MSE) loss for training our network. Instead of using Perceptual Loss, we utilize LPIPS loss for training. Given the ground-truth target view of dimension $\hat{I} \in \mathbb{R}^{H \times W \times C}$ and the reconstructed target view $I$, the effective objective function used for optimization is defined as:

\begin{equation} L = MSE(I,\hat{I}) + \lambda \cdot LPIPS(I, \hat{I}) \end{equation}

where $\lambda$ is a scaling factor set to $0.5$ for all experiments.

\subsection{Emergent Properties}

One surprising \textbf{emergent property} of our newly proposed transformer block is its ability to disentangle the source and target tokens, which allows it to scale better for synthetic data compared to a naive transformer block. We present these results in Figure X, where we observe significant improvements. We hypothesize that this emergent property arises because synthetic data is generally prone to artifacts and out-of-distribution noise. When transformer blocks cannot distinguish between source and target tokens, the model learns using both real and synthetic data, leading to reconstructions that inherit these artifacts. However, in our case, only the relevant information from clean images is used for backpropagation, allowing the model to utilize useful context from synthetic data while discarding artifacts during token fusion for target view generation.

\section{Limitations}

\begin{figure*}
    \centering
    \includegraphics[width=\textwidth]{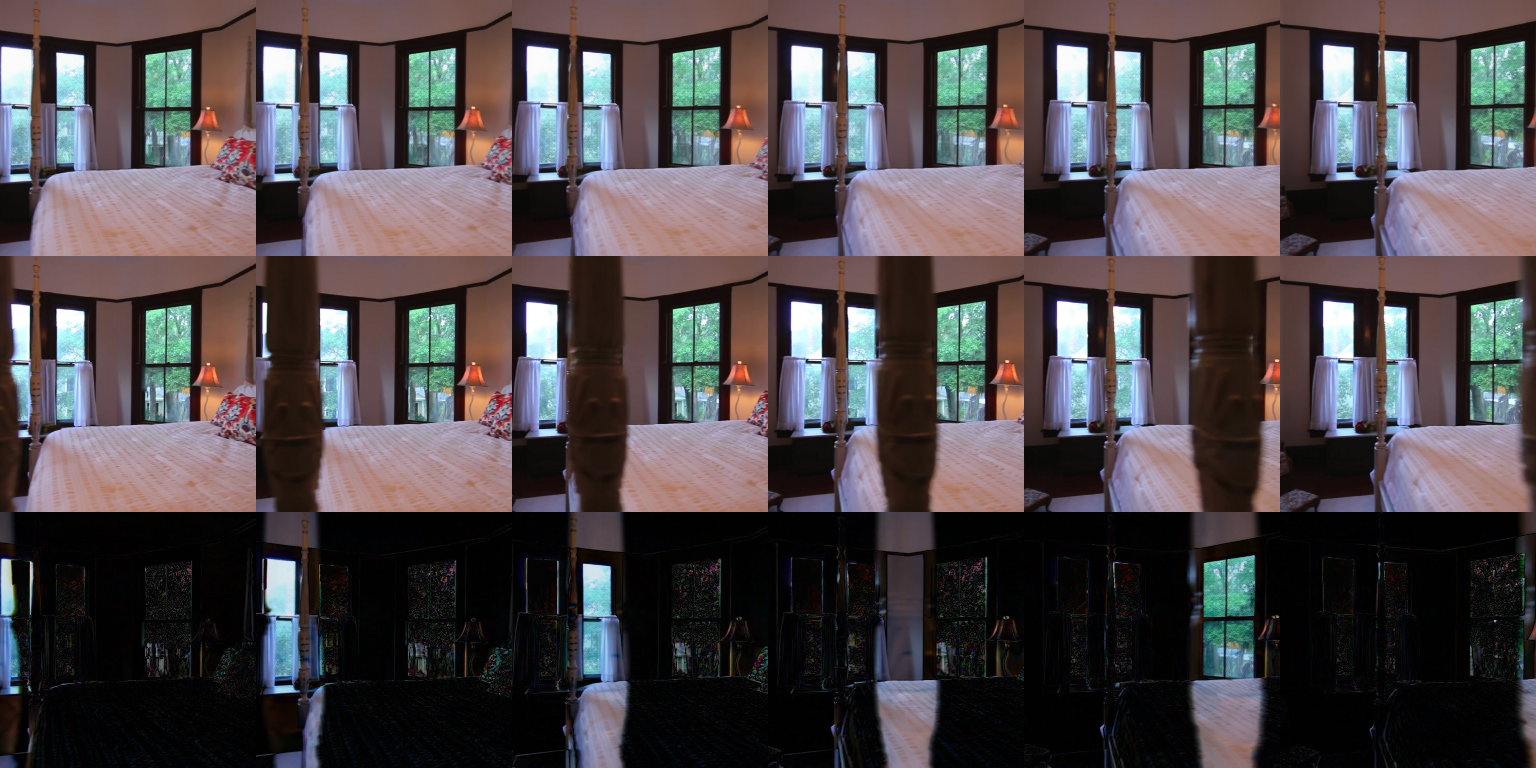}
    \caption{\textbf{Figure illustrating failure cases of our method}. Our method fails to perform well if there are occluded objects coming into the scene. Figure ordering is OURS, GT, DIFF}
    \label{fig:limitation}
\end{figure*}

Our model struggles when regions occluded in the source images become visible in the target view. As shown in \Cref{fig:limitation}, when a new object enters the scene, the model hallucinates the affected region. We argue that this phenomenon is inherently ill-posed and lacks a definitive solution. Additionally, the model uses a token size of 8 for all blocks, resulting in 1024 tokens per source image, which demands significant memory. We leave further architectural optimizations, such as hierarchical transformers and more efficient networks like linear attention and state-space models (e.g., Mamba \cite{gu2024mambalineartimesequencemodeling}, \cite{zhu2024visionmambaefficientvisual}), for future work.

\begin{figure*}[htbp]
    \centering
    \captionsetup[subfigure]{aboveskip=0pt, belowskip=0pt}
    \subfloat{\includegraphics[width=\textwidth]{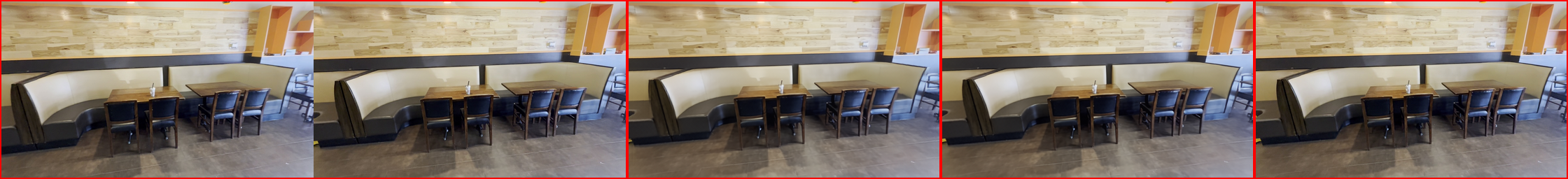}}\vspace{-3mm}
    \subfloat{\includegraphics[width=\textwidth]{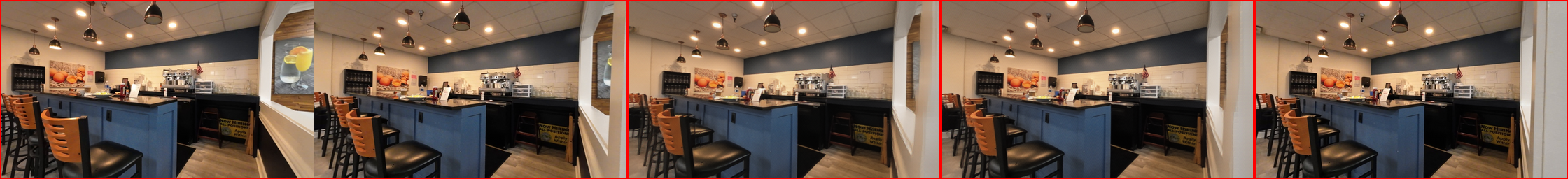}}\vspace{-3mm}
    \subfloat{\includegraphics[width=\textwidth]{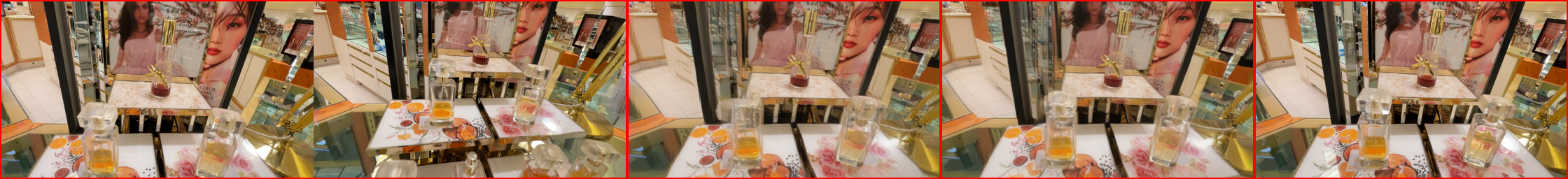}}\vspace{-3mm}
    \subfloat{\includegraphics[width=\textwidth]{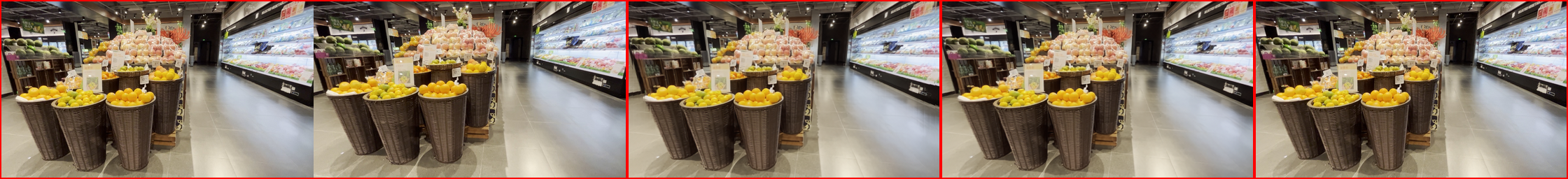}}\vspace{-3mm}
    \subfloat{\includegraphics[width=\textwidth]{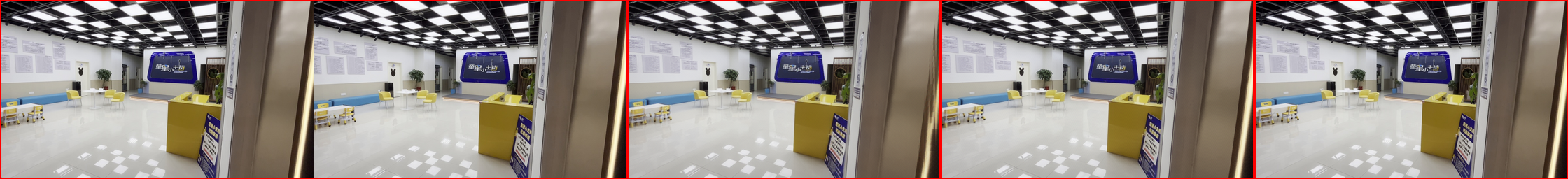}}\vspace{-3mm}
    \subfloat{\includegraphics[width=\textwidth]{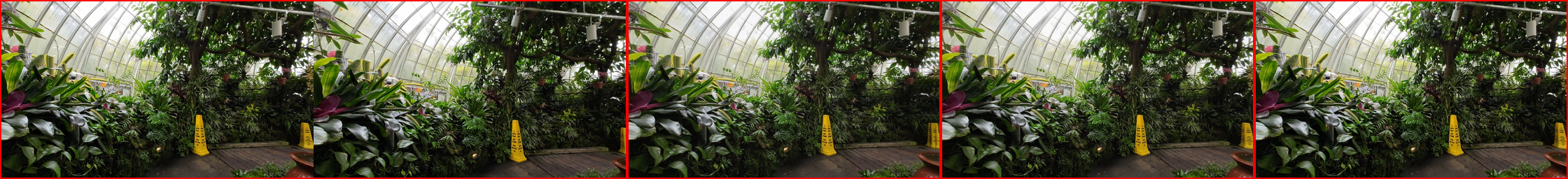}}\vspace{-3mm}
        \caption{\textbf{Figure illustrating results from DL3DV dataset trained with our synthetic data}. The first 2 images represent the input views. third presents results of LVSM, Fourth represents our results and fifth the ground truth}
    \label{fig:compact_subdl3dv}
\end{figure*}

\begin{figure*}[htbp]
    \centering
    \captionsetup[subfigure]{aboveskip=0pt, belowskip=0pt}
    \subfloat{\includegraphics[width=\textwidth]{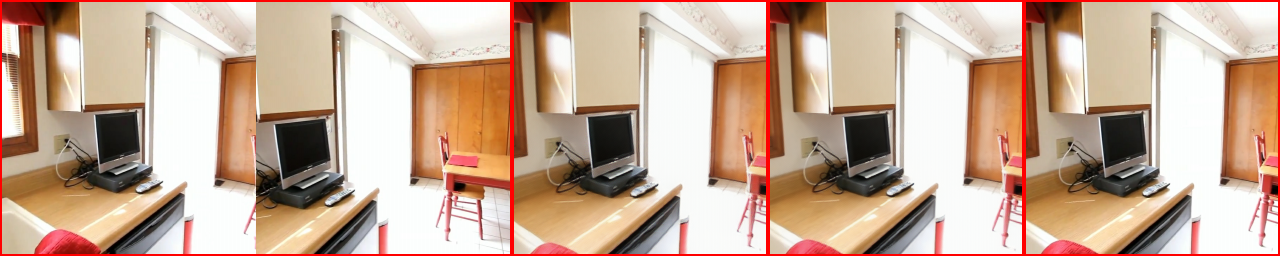}}\vspace{-3mm}
    \subfloat{\includegraphics[width=\textwidth]{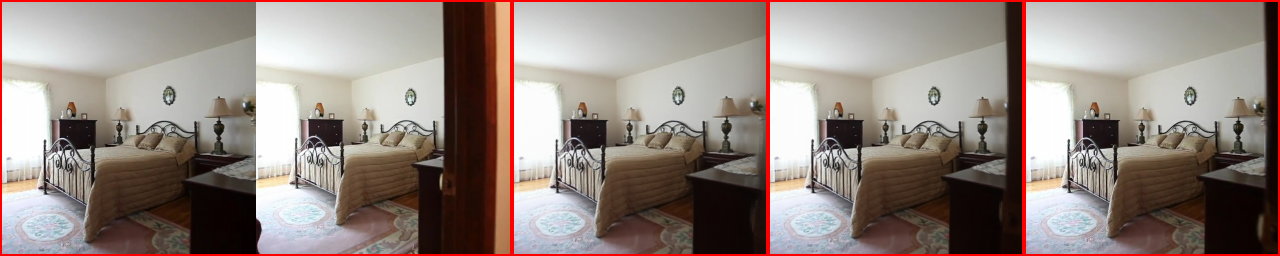}}\vspace{-3mm}
    \subfloat{\includegraphics[width=\textwidth]{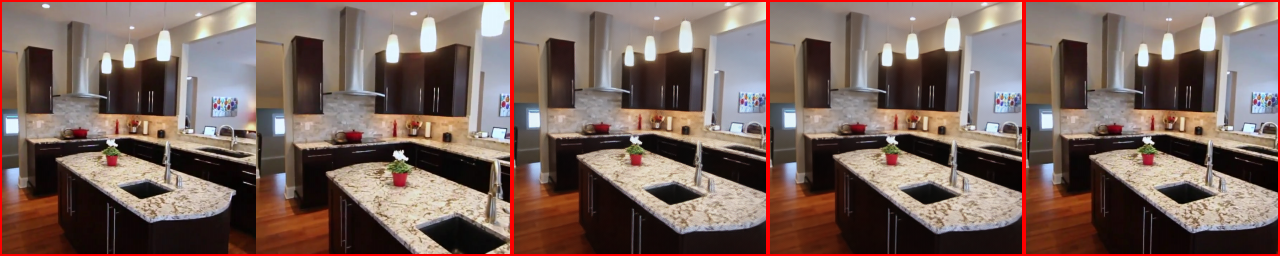}}\vspace{-3mm}
    \subfloat{\includegraphics[width=\textwidth]{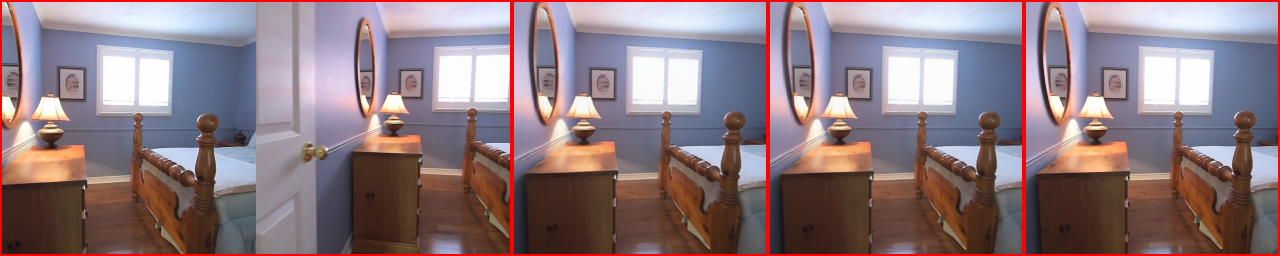}}\vspace{-3mm}
    \subfloat{\includegraphics[width=\textwidth]{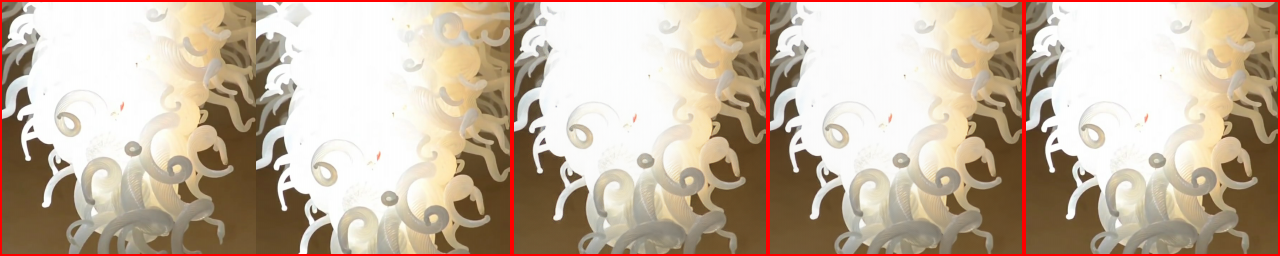}}\vspace{-3mm}
    \subfloat{\includegraphics[width=\textwidth]{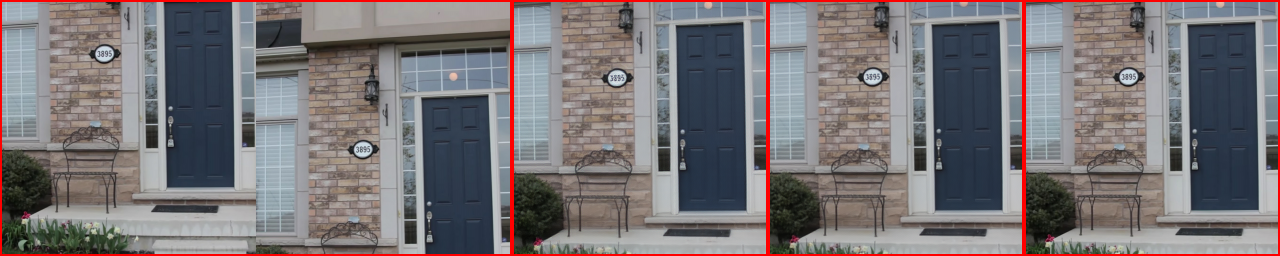}}\vspace{-3mm}
        \caption{\textbf{Figure illustrating results from Re10k dataset trained with our synthetic data}. The first 2 images represent the input views. third presents results of LVSM, Fourth represents our results and fifth the ground truth}
    \label{fig:compact_subre10k}
\end{figure*}

\begin{figure*}[htbp]
    \centering
    \captionsetup[subfigure]{aboveskip=0pt, belowskip=0pt}
    \subfloat{\includegraphics[width=\textwidth]{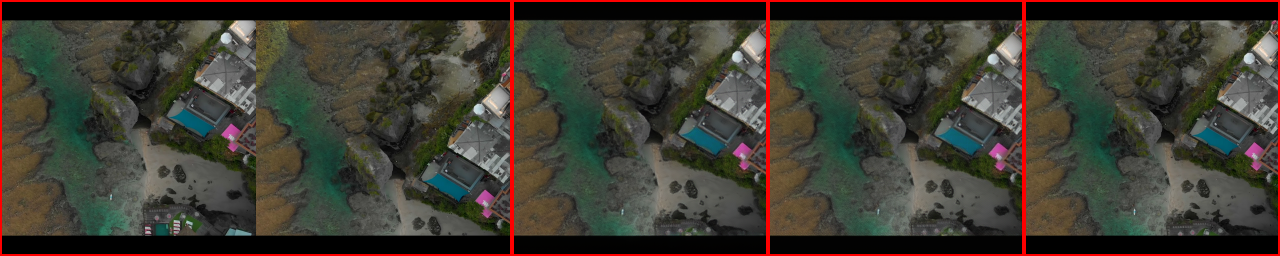}}\vspace{-3mm}
    \subfloat{\includegraphics[width=\textwidth]{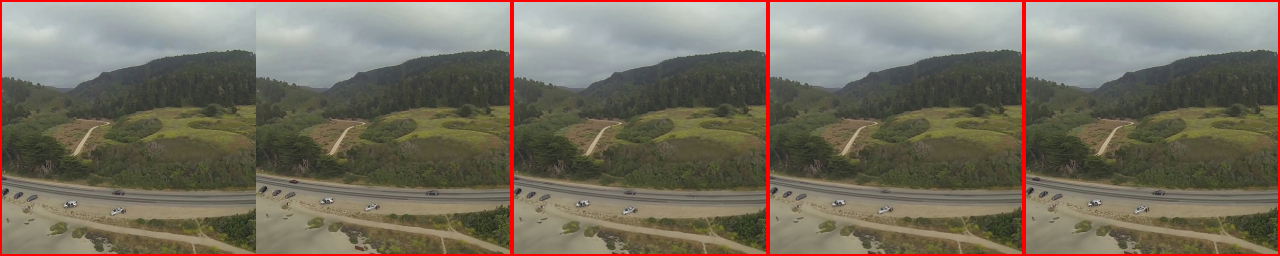}}\vspace{-3mm}
    \subfloat{\includegraphics[width=\textwidth]{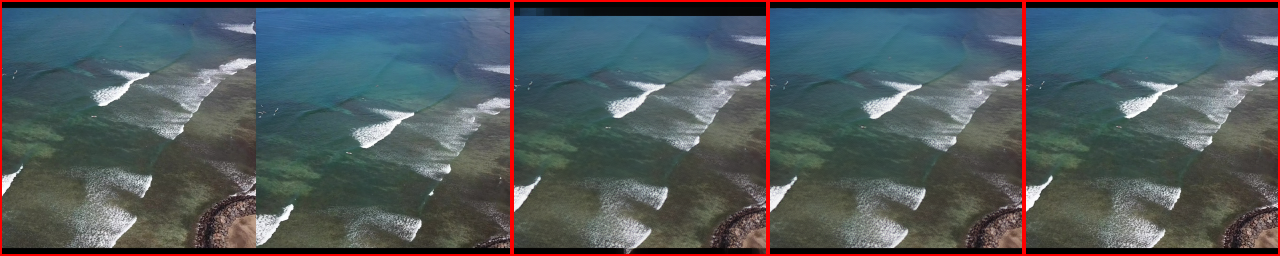}}\vspace{-3mm}
    \subfloat{\includegraphics[width=\textwidth]{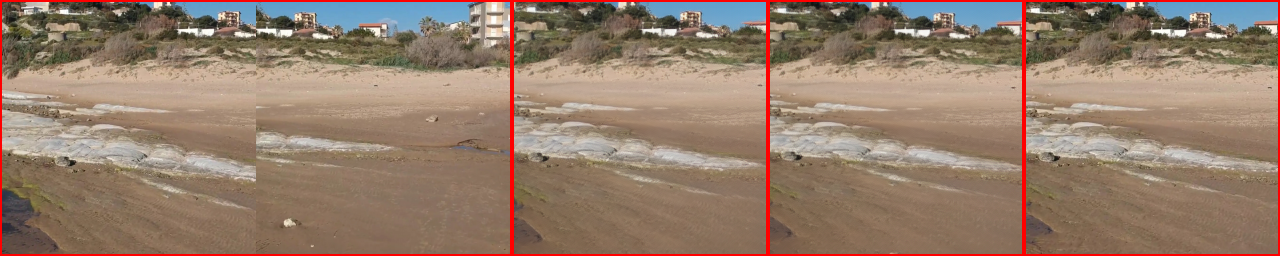}}\vspace{-3mm}
    \subfloat{\includegraphics[width=\textwidth]{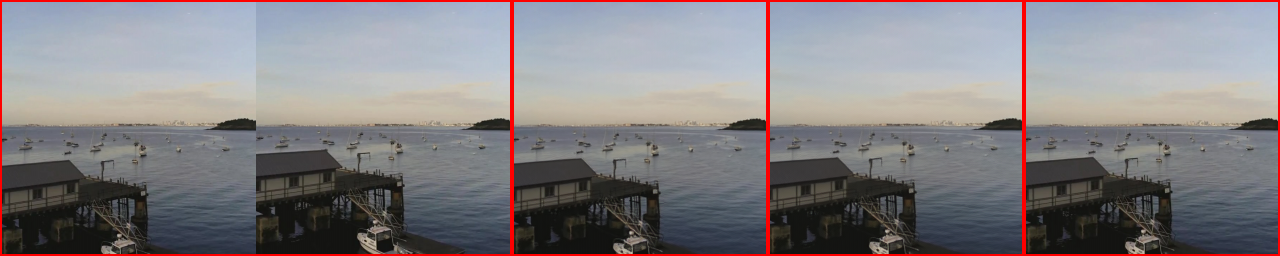}}\vspace{-3mm}
    \caption{\textbf{Figure illustrating results from ACID dataset trained with our synthetic data}. The first 2 images represent the input views. third presents results of LVSM, Fourth represents our results and fifth the ground truth}
    \label{fig:compact_subacid}
\end{figure*}

\begin{figure*}
    \centering
    \includegraphics[width=0.8\textwidth]{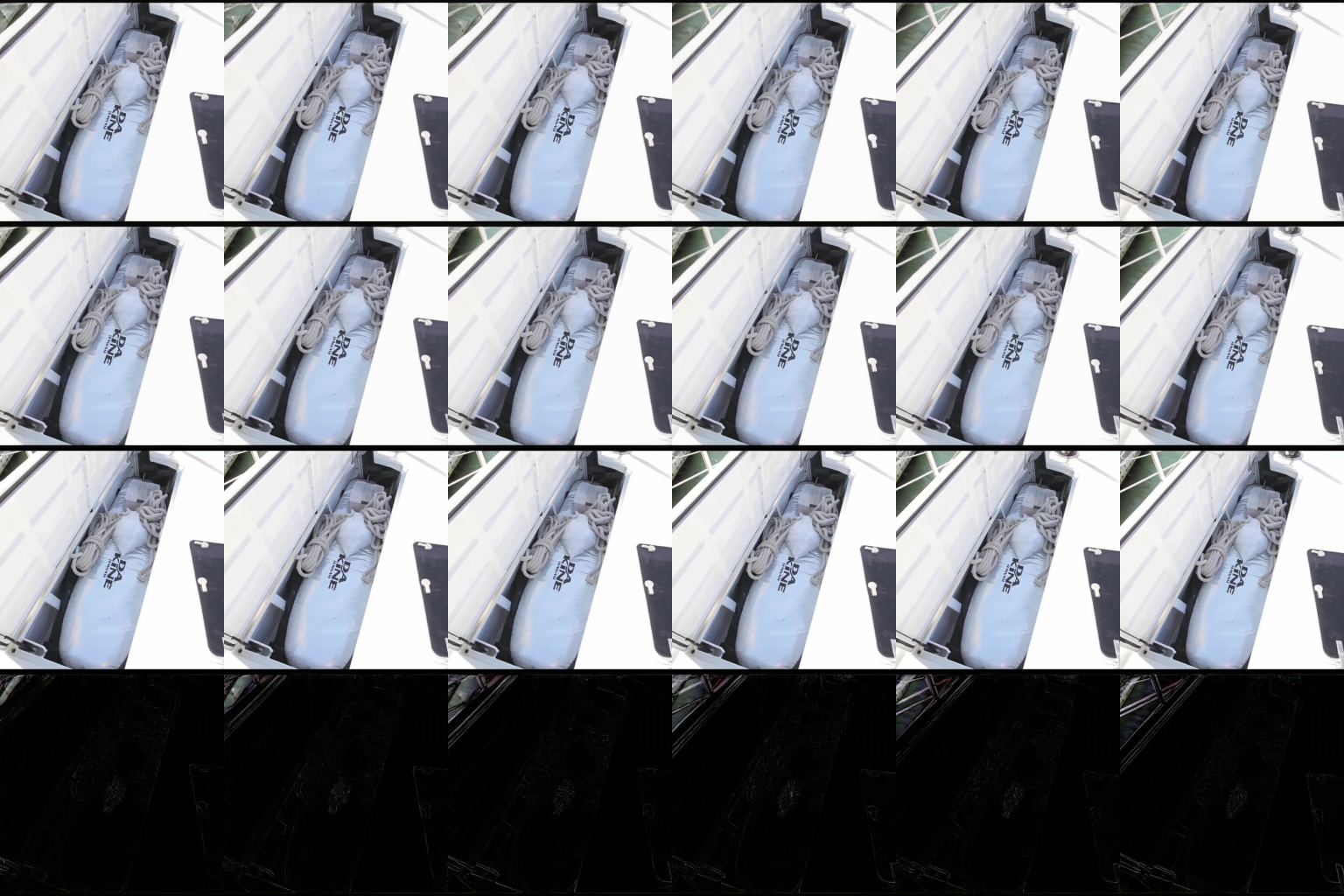}
    \caption{ \textbf{Figure illustrating the regions where our method works better than LVSM for Re10K dataset}. The figures are in the order Row 1:- LVSM, Row 2:- OURS Row 3:- GT, Row 4:- Difference between LVSM and Ours}
    \label{fig:my_label1}
\end{figure*}

\begin{figure*}
    \centering
    \includegraphics[width=0.8\textwidth]{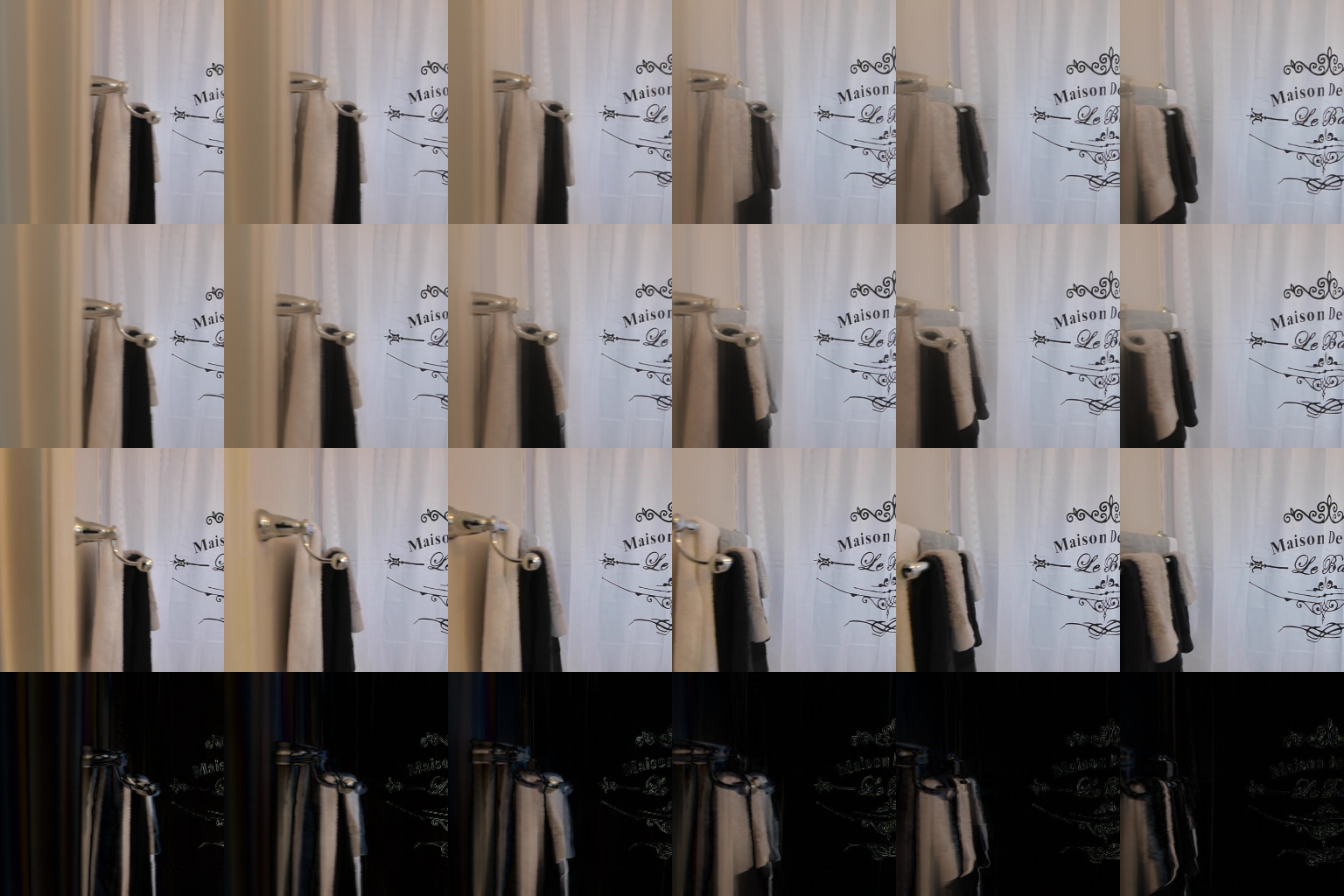}
    \caption{ \textbf{Figure illustrating the regions where our method works better than LVSM for Re10K dataset}. The figures are in the order Row 1:- LVSM, Row 2:- OURS Row 3:- GT, Row 4:- Difference between LVSM and Ours}
    \label{fig:my_label1}
\end{figure*}

\begin{figure*}
    \centering
    \includegraphics[width=0.8\textwidth]{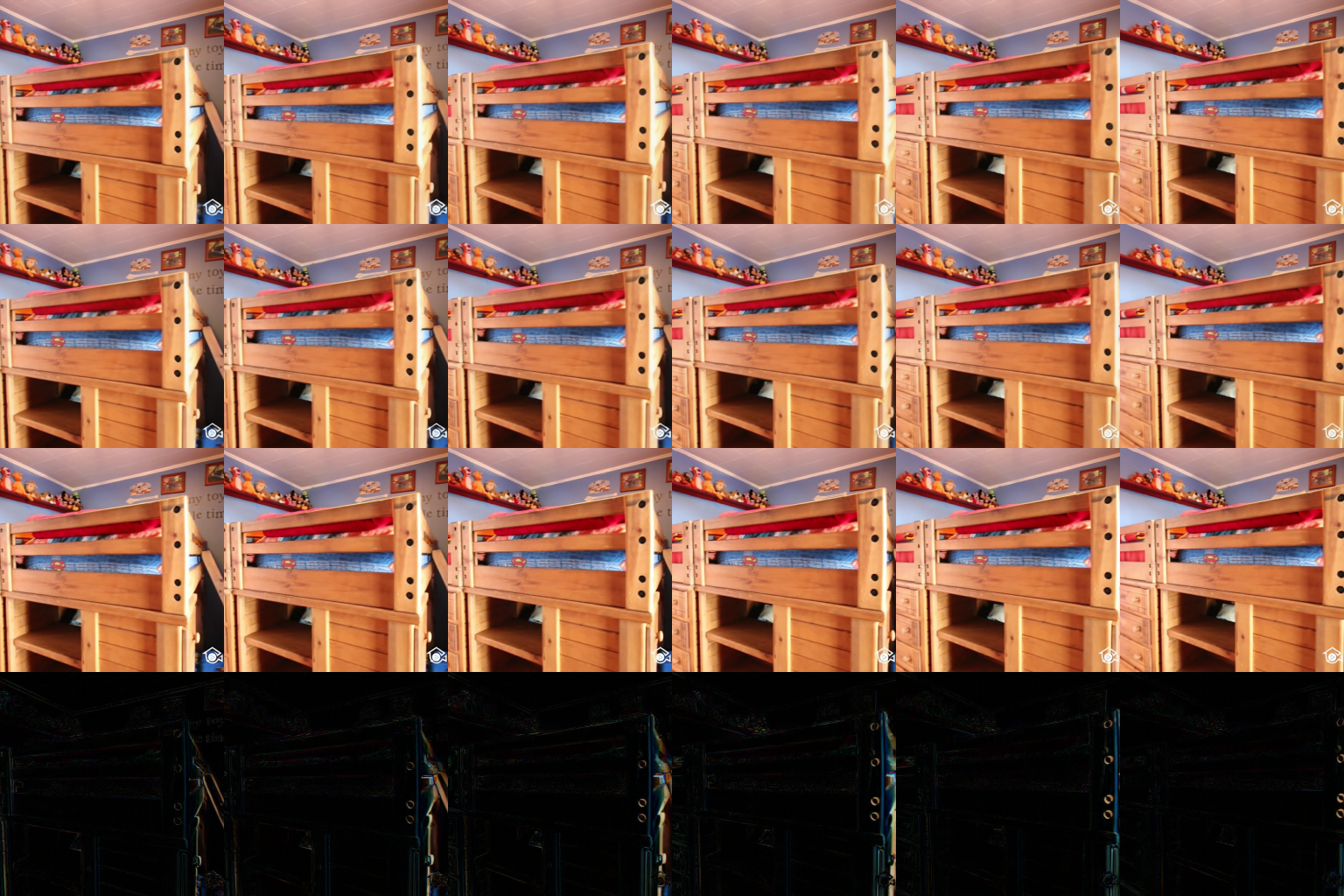}
    \caption{ \textbf{Figure illustrating the regions where our method works better than LVSM for Re10K dataset}. The figures are in the order Row 1:- LVSM, Row 2:- OURS Row 3:- GT, Row 4:- Difference between LVSM and Ours}
    \label{fig:my_label1}
\end{figure*}

\begin{figure*}
    \centering
    \includegraphics[width=0.8\textwidth]{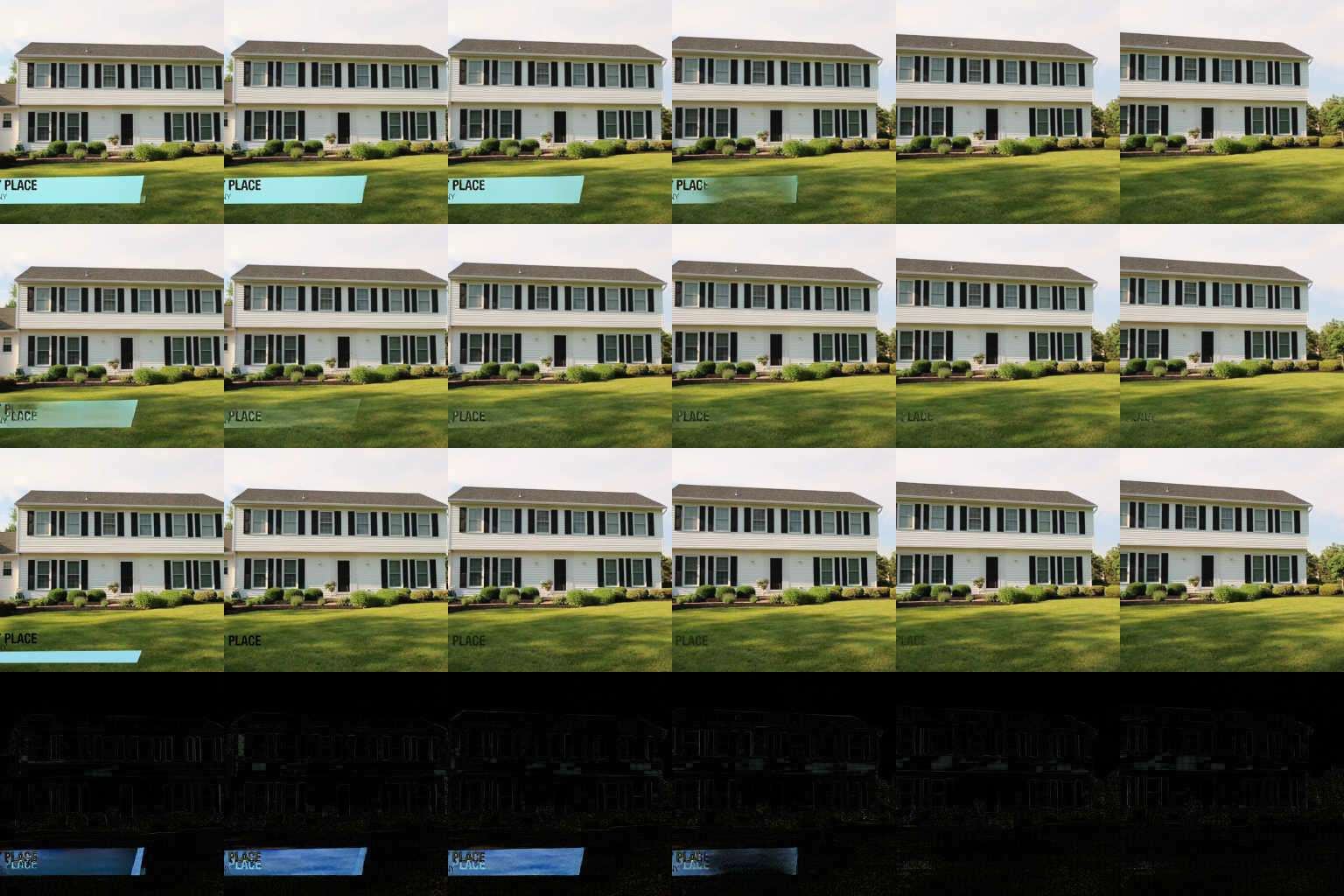}
    \caption{ \textbf{Figure illustrating the regions where our method works better than LVSM for Re10K dataset}. The figures are in the order Row 1:- LVSM, Row 2:- OURS Row 3:- GT, Row 4:- Difference between LVSM and Ours}
    \label{fig:my_label1}
\end{figure*}

\begin{figure*}
    \centering
    \includegraphics[width=0.8\textwidth]{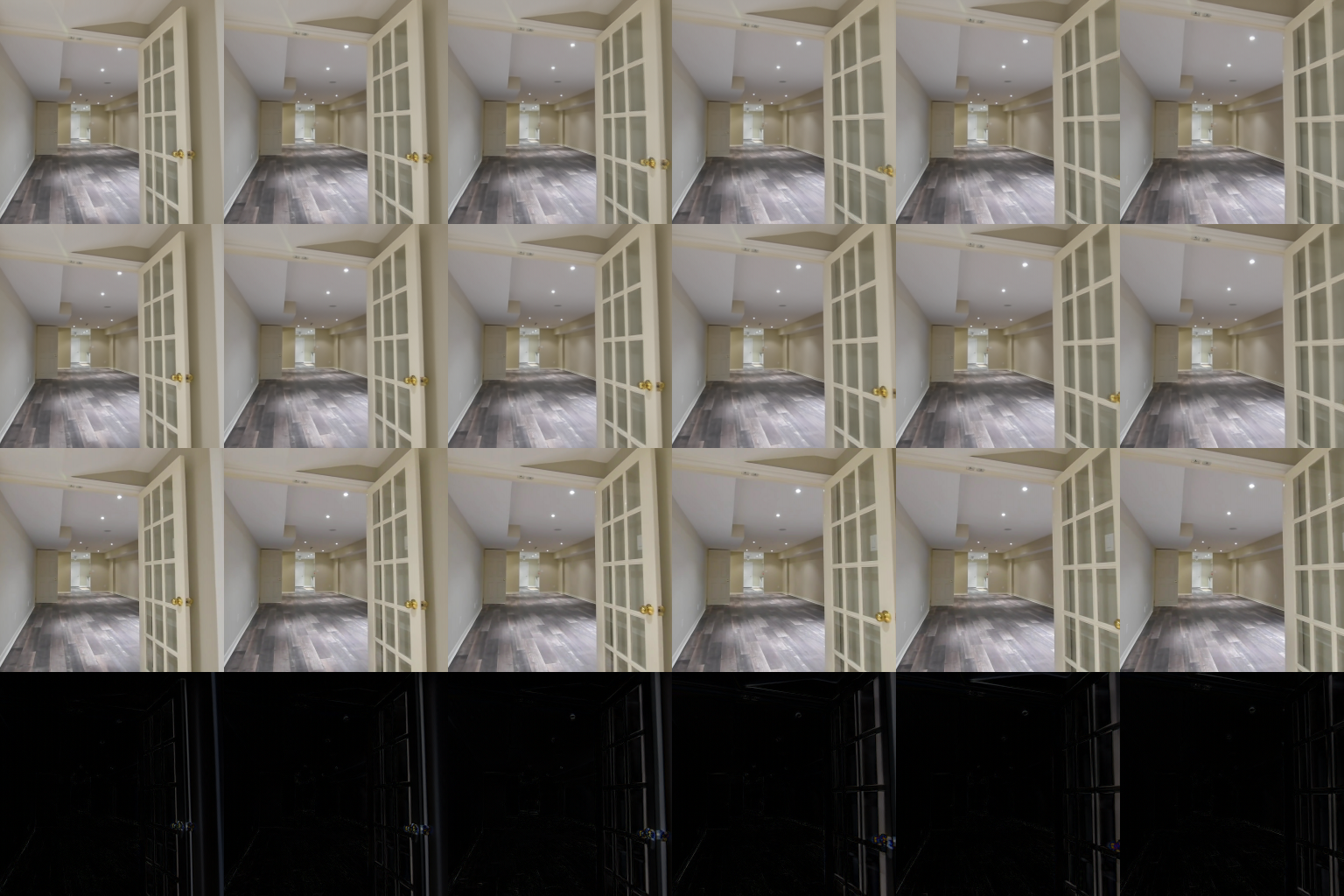}
    \caption{ \textbf{Figure illustrating the regions where our method works better than LVSM for Re10K dataset}. The figures are in the order Row 1:- LVSM, Row 2:- OURS Row 3:- GT, Row 4:- Difference between LVSM and Ours}
    \label{fig:my_label1}
\end{figure*}

\begin{figure*}
    \centering
    \includegraphics[width=0.8\textwidth]{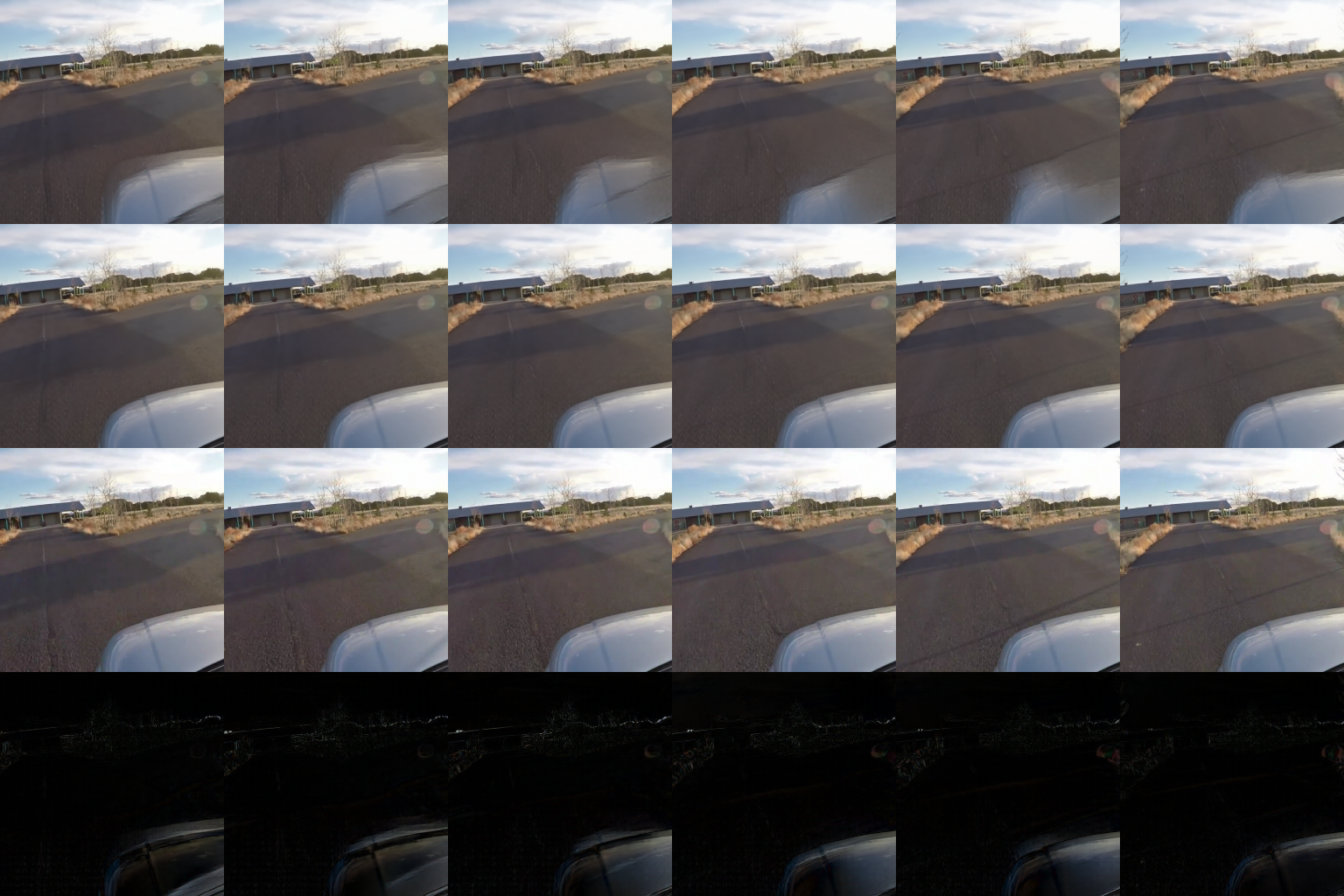}
    \caption{ \textbf{Figure illustrating the regions where our method works better than LVSM for Re10K dataset}. The figures are in the order Row 1:- LVSM, Row 2:- OURS Row 3:- GT, Row 4:- Difference between LVSM and }
    \label{fig:my_label1}
\end{figure*}

\begin{figure*}
    \centering
    \includegraphics[width=0.8\textwidth]{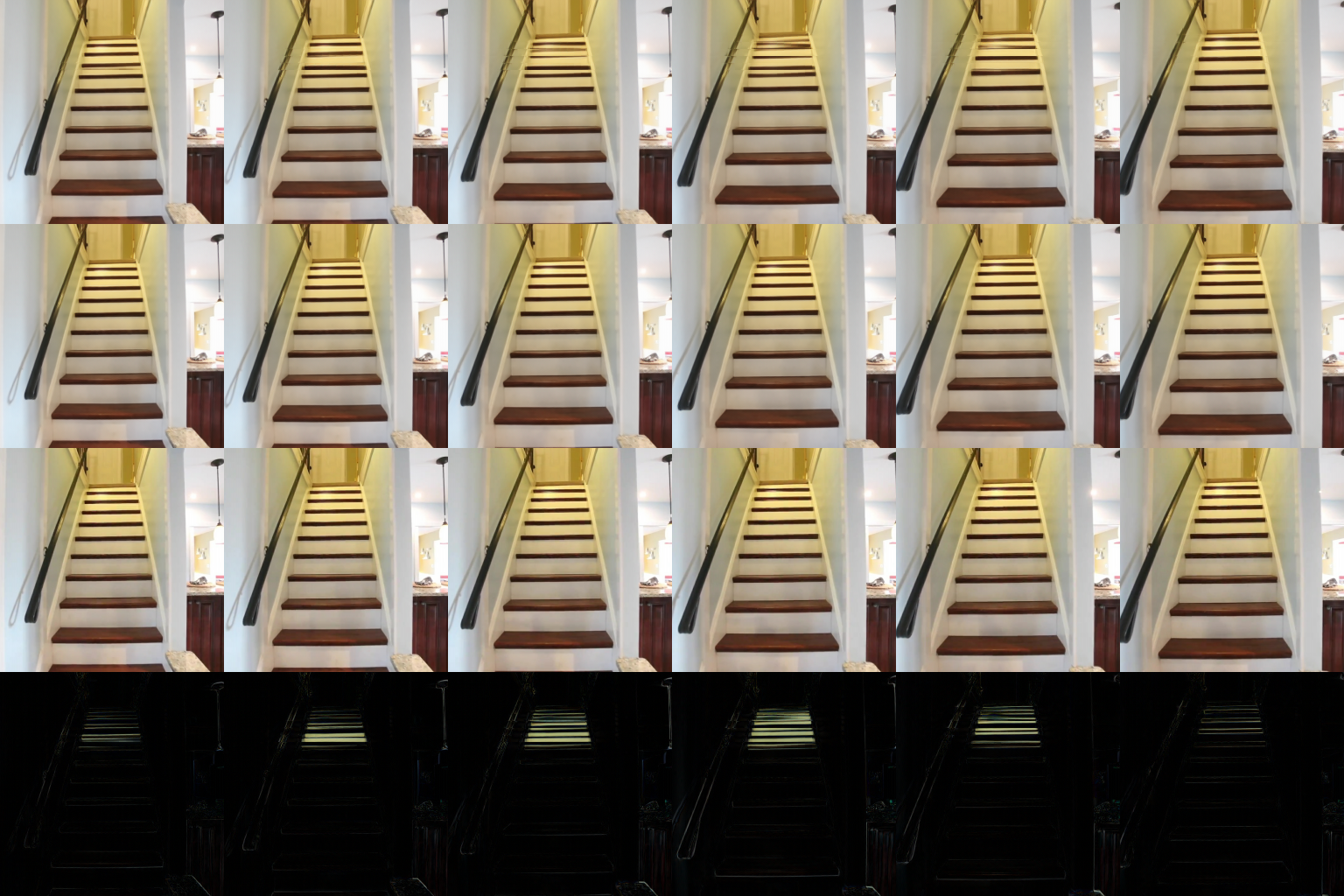}
    \caption{ \textbf{Figure illustrating the regions where our method works better than LVSM for Re10K dataset}. The figures are in the order Row 1:- LVSM, Row 2:- OURS Row 3:- GT, Row 4:- Difference between LVSM and Ours}
    \label{fig:my_label1}
\end{figure*}

\section{Failure cases of our method}

We notice that our method contains two main failure modes (1) when an new object comes into the view in between the conditioned frames. (2) When too many shiny artifacts are present in the image

\begin{figure*}
    \centering
    \includegraphics[width=\textwidth]{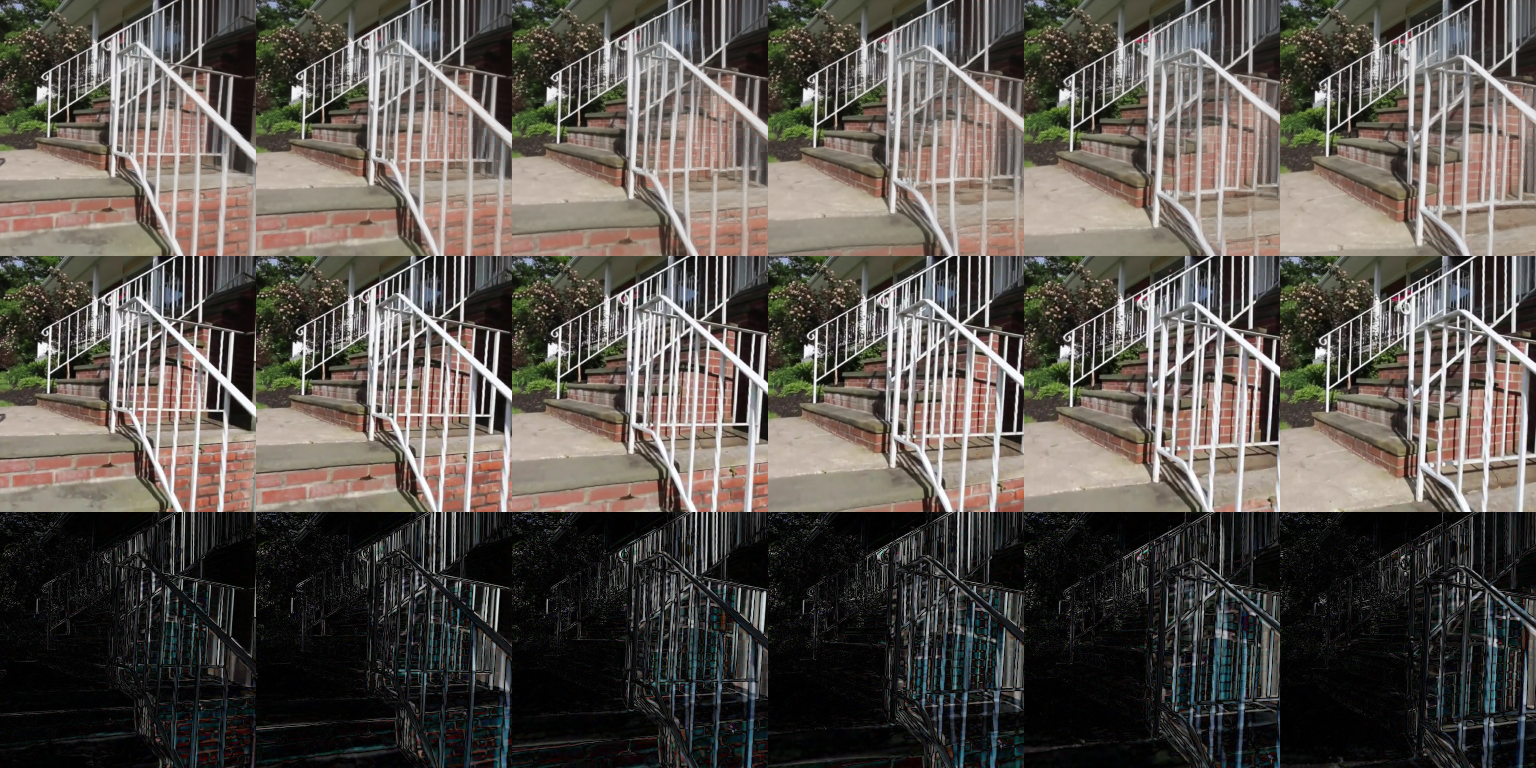}
    \caption{\textbf{Figure illustrating failure cases of our method}. Our method fails to perform well if there are occluded objects coming into the scene. Figure ordering is OURS, GT, DIFF}
    \label{fig:fail4}
\end{figure*}

\begin{figure*}
    \centering
    \includegraphics[width=\textwidth]{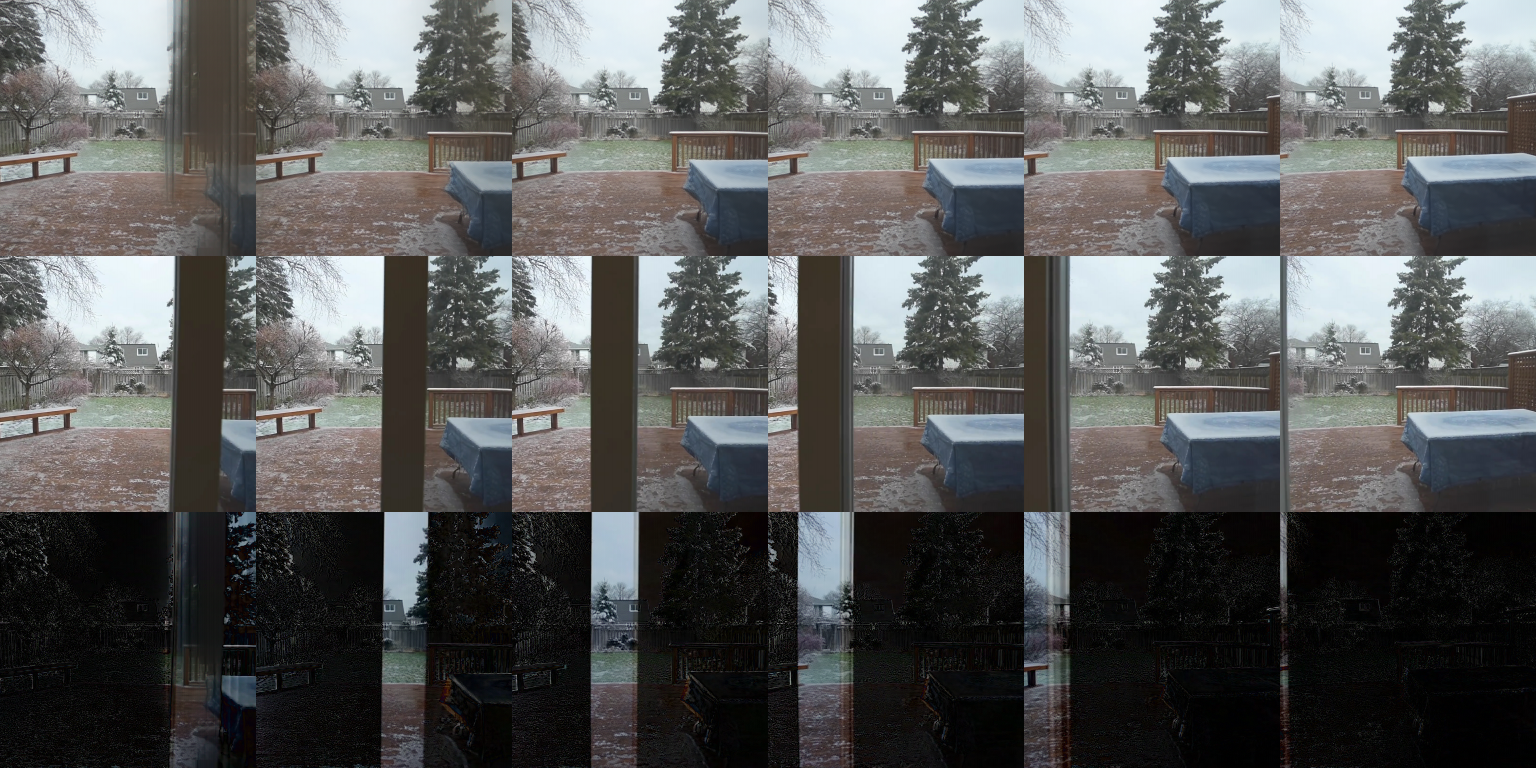}
    \caption{\textbf{Figure illustrating failure cases of our method}. Our method fails to perform well if there are occluded objects coming into the scene. Figure ordering is OURS, GT, DIFF}
    \label{fig:fail3}
\end{figure*}

\begin{figure*}
    \centering
    \includegraphics[width=\textwidth]{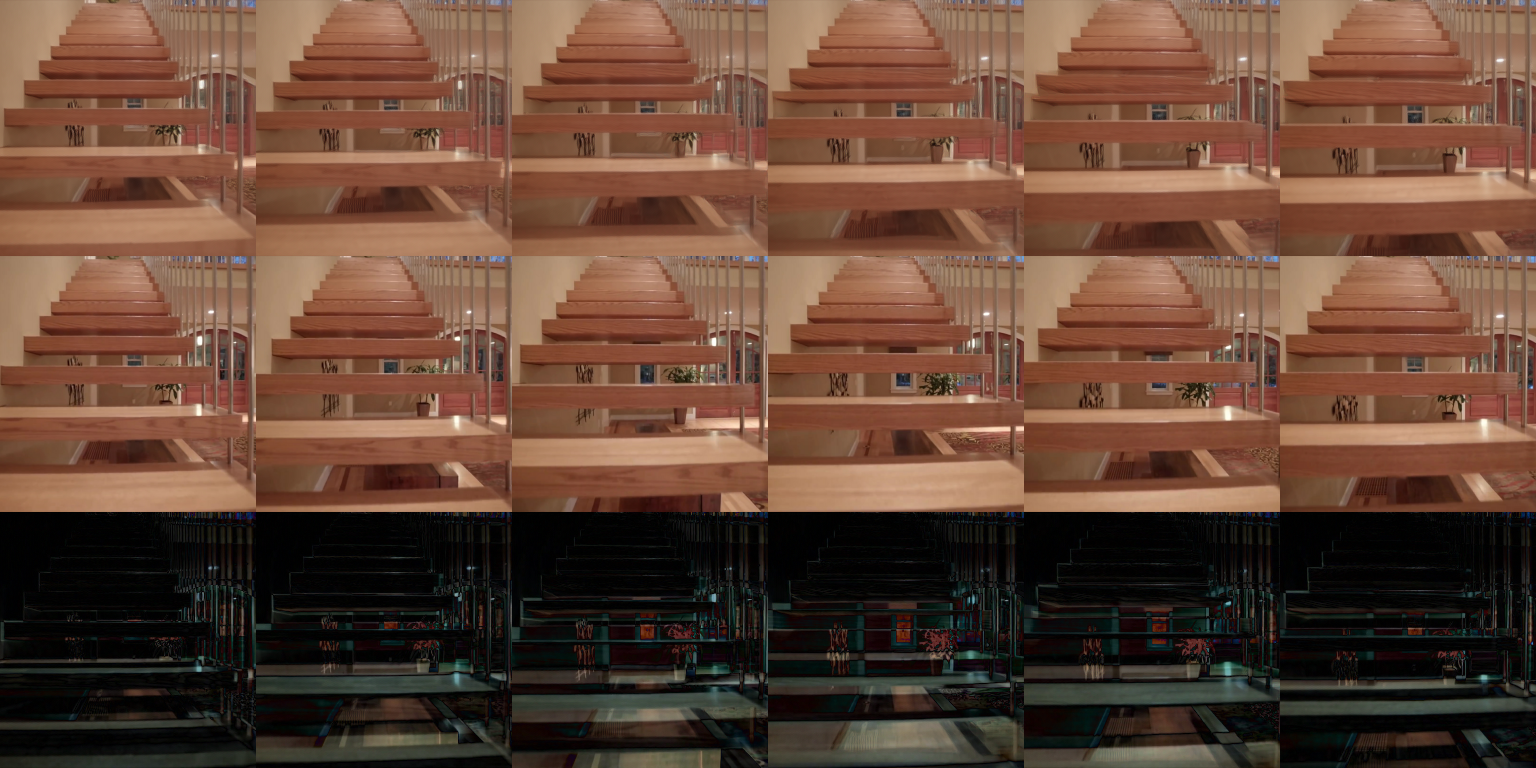}
    \caption{\textbf{Figure illustrating failure cases of our method}. Our method fails to perform well if there are occluded objects coming into the scene. Figure ordering is OURS, GT, DIFF}
    \label{fig:fail2}
\end{figure*}

\begin{figure*}
    \centering
    \includegraphics[width=\textwidth]{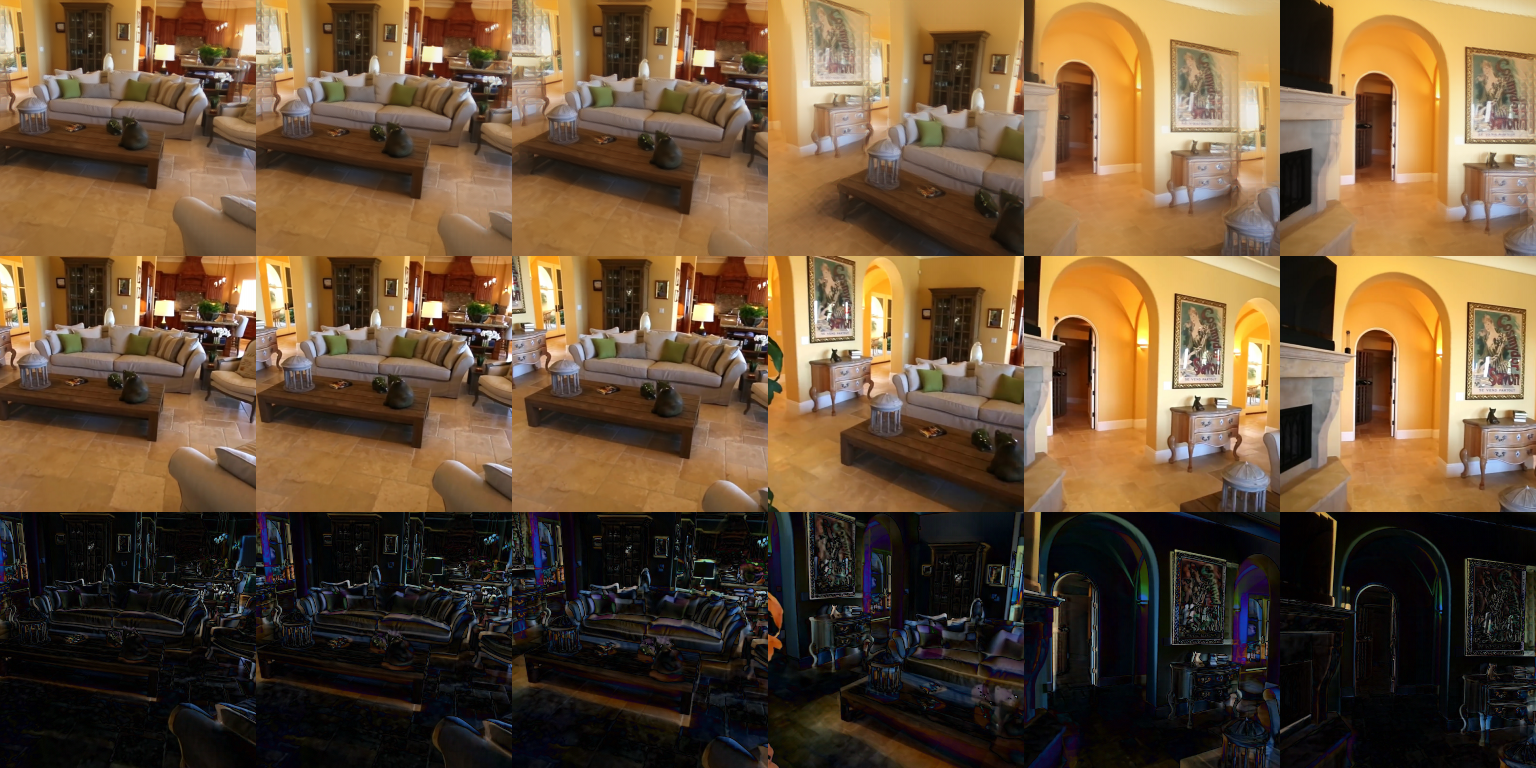}
    \caption{\textbf{Figure illustrating failure cases of our method}. Our method fails to perform well if there are occluded objects coming into the scene moreover, our method also fails to reconstruct properly when there are some shiny obejcts in the scene. Figure ordering is OURS, GT, DIFF}
    \label{fig:fail1}
\end{figure*}

\end{document}